\documentclass[aps,twocolumn,nofootinbib,groupedaddress,amsfonts,floatfix
]{revtex4-1} 
\usepackage{graphicx,amsmath,amssymb,amstext}
\usepackage{amssymb,amsbsy,amsfonts,amsthm,color}
\usepackage{comment}
\usepackage{cancel}

\usepackage{mathtools,nccmath}

\usepackage{epsfig}
\usepackage{graphicx}
\usepackage{subfigure}
\usepackage[dvipsnames]{xcolor}
\definecolor{linkcolor}{rgb}{0.0,0.3,0.5}
\usepackage[unicode, colorlinks=true, linkcolor=linkcolor, citecolor=linkcolor, 
filecolor=linkcolor,urlcolor=linkcolor, pdfusetitle]{hyperref}
\usepackage{cancel}
\usepackage{balance}
\usepackage[capitalise]{cleveref}
\usepackage{xspace}
\usepackage{enumerate}
\usepackage{ulem}
\usepackage{mdframed}
\usepackage{booktabs}

\usepackage{epigraph}

\AtBeginDocument{%
  \heavyrulewidth=.08em
  \lightrulewidth=.05em
  \cmidrulewidth=.03em
  \belowrulesep=.65ex
  \belowbottomsep=0pt
  \aboverulesep=.4ex
  \abovetopsep=0pt
  \cmidrulesep=\doublerulesep
  \cmidrulekern=.5em
  \defaultaddspace=.5em
}

\graphicspath{{Figures/}}

\usepackage{color}

\newcommand{\Beq}{\begin{eqnarray}}
\newcommand{\Eeq}{\end{eqnarray}}

\newcommand{\nn}{\nonumber \\}

\def\lsim{\mathrel {\vcenter {\baselineskip 0pt \kern 0pt \hbox{$<$} \kern 0pt \hbox{$\sim$} }}}

\def\gsim{\mathrel {\vcenter {\baselineskip 0pt \kern 0pt \hbox{$>$} \kern 0pt \hbox{$\sim$} }}}

\newcommand{\RomanNumeralCaps}[1]

\def\-{\,-\,}
\def\={\,=\,}
\def\+{\,+\,}
\def\equi{\,\equiv\,}

\definecolor{mypurple}{RGB}{143, 116, 210}

\newcommand{\osu}{Department of Physics and Center for Cosmology and AstroParticle Physics (CCAPP),\\
The Ohio State University, Columbus, OH 43210, USA}

\begin{document}
{\hfill 
\title{Tidal Disruption in Topological Solitons \\ and the Emergence of an Effective Horizon}

\author{Pierre Heidmann}
\affiliation{\osu}

\author{Gela Patashuri $\,$}
\email{heidmann.5@osu.edu; patashuri.1@osu.edu}
\affiliation{\osu}

\begin{abstract}
We compute the dynamics of particles and strings falling into smooth horizonless spacetimes that match the Schwarzschild black hole but replace its horizon with a smooth cap in supergravity. The cap consists of a regular topological structure formed by the deformations of extra compact dimensions. We show that infalling particles follow Schwarzschild-like trajectories down to the cap, but experience rapidly growing tidal forces that reach extreme values. In addition, infalling strings encounter a region of tidal instability localized at the cap, where transverse modes are excited. This stringy excitation drains their kinetic energy, resulting in tidal trapping. We demonstrate that the onset and strength of this instability depend sensitively on the Kaluza-Klein scale, the string scale, and the mass of the spacetime, ensuring that strings cannot escape the cap region. These results show that horizonless geometries can reproduce key features of black hole absorption while maintaining regularity at the horizon scale, offering compelling evidence for the emergence of effective horizon-like behavior from topological spacetime structures.
\end{abstract}
\maketitle


\section{Introduction} \label{sect:intro2}

Recent advances have led to a powerful solution-generating technique that enables the construction of non-supersymmetric, ultracompact, and horizonless geometries in supergravity \cite{Heidmann:2021cms,*Bah:2020pdz,*Bah:2021owp,*Bah:2021rki,*Heidmann:2022zyd,*Bah:2022pdn,Chakraborty:2025ger}. These geometries replace the horizons of nonextremal black holes with smooth, regular ``caps'' formed by Kaluza-Klein (KK) bubbles and supported by electromagnetic fluxes. Known as \emph{topological solitons}, these solutions provide explicit, controlled prototypes of quantum microstates of nonextremal black holes in string theory. Because they are sufficiently coherent to admit a classical, smooth spacetime, they offer a rare and valuable window into the horizon-scale microstructure underlying nonextremal black holes. 

The procedure has been successfully applied to the most iconic black hole in General Relativity (GR): the Schwarzschild black hole. In \cite{Bah:2022yji,Bah:2023ows}, Schwarzschild topological solitons were constructed and analyzed. More recently, in \cite{Heidmann:2023kry}, regular and neutral bound states of oppositely charged extremal black holes, stabilized by a KK bubble, have been derived. While the latter solutions are not entirely horizonless, they possess key advantages over the earlier topological solitons. First, their simpler structure allows for a more tractable analysis. Second, the Schwarzschild horizon is effectively resolved: the classical $r=2\mathcal{M}$ surface is replaced by a smooth KK bubble, with the extremal black holes being confined at its poles, so that some physical probes interact only with the KK bubble. Third, these configurations account for a substantial fraction of the Schwarzschild entropy, rendering them promising candidates for capturing the quantum-gravitational physics that emerges at the horizon scale of the Schwarzschild black hole.

In this paper, we use these explicit geometries to address a fundamental question in the understanding of regular black-hole microstructure in string theory:
How can a horizonless topological geometry reproduce the absorptive behavior of a classical black hole? More precisely, \emph{how can an effective horizon emerge from a geometry that is everywhere smooth and horizonless?} 

In classical GR, absorption follows trivially from the presence of a horizon. By contrast, a probe falling into a horizonless geometry with a smooth topological cap would seemingly bounce off and escape to infinity. A black hole-like absorption in such a background must therefore emerge dynamically, through nontrivial interactions between the geometry and the probe internal structure. But how can a geometry, ending in a smooth cap and nearly identical to Schwarzschild up to the horizon scale, induce such irreversible dynamics on \emph{any} probes?

\begin{figure*}[t]
    \begin{center}
    \includegraphics[width=0.76\linewidth]{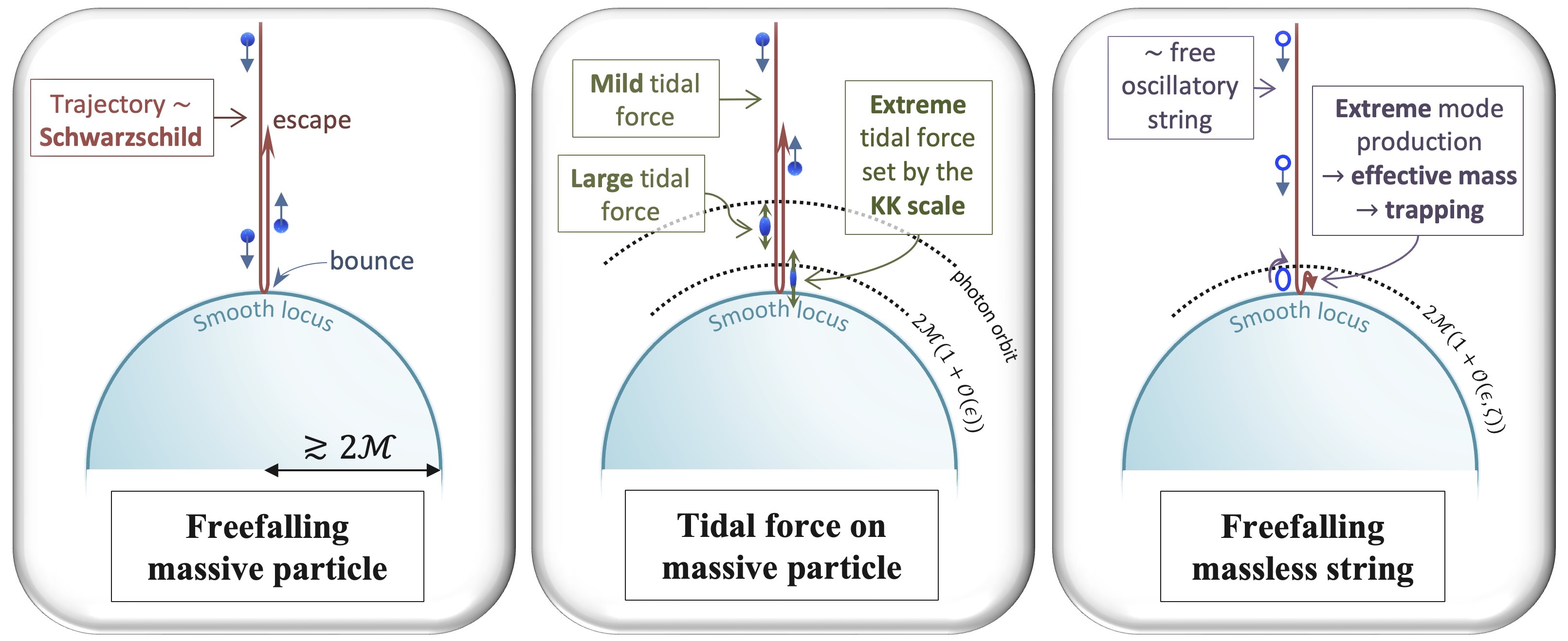}
    \caption{\textit{Schematic illustration of how smooth, horizonless Schwarzschild topological solitons exhibit absorptive, horizon-like behavior. Infalling point particles follow nearly Schwarzschild geodesics, but experience extreme tidal forces near $r \sim 2\mathcal{M}$. These tidal effects trigger internal string excitations, leading to effective trapping at the cap.}}
    \label{fig:Intro}
    \end{center}
\end{figure*}

This question has been addressed in the context of supersymmetric smooth geometries in a series of studies \cite{Tyukov:2017uig,Bena:2018mpb,Bena:2020iyw,Martinec:2020cml}. Despite their special and unrealistic nature, supersymmetric solutions provide valuable insights into how smooth geometries might mimic black hole dynamics. These works demonstrated that infalling point particles experience extreme tidal stresses well before reaching the cap. When the probe has internal degrees of freedom, such as a string, the tidal forces excite internal modes and drain kinetic energy, effectively trapping the object within the geometry. This mechanism was made precise in \cite{Martinec:2020cml}, where it was shown that infalling strings undergo tidal excitation and instability, gain effective mass, and are dynamically trapped after bouncing off the cap.

However, an atypicality of the supersymmetric case is that the tidal excitation and trapping occur significantly outside the cap, well above the expected scale where new quantum gravitational effects should arise. In the present paper, we demonstrate how more realistic Schwarzschild topological solitons correct this issue by confining new physics closer to the would-be horizon: what we refer to as the horizon scale. We do so by analyzing the dynamics of massive point particles and massless strings in free fall within these geometries.

We will show that the essential mechanism (tidal amplification and trapping) persists, but with key improvements: all extreme tidal effects are confined near the smooth cap at the horizon scale,
\begin{equation}
    r\,\sim\, 2\mathcal{M}\left( 1+ \mathcal{O}(\epsilon)\right),
\end{equation}
where $\mathcal{M}$ is the ADM mass and $\epsilon = \left(R_\psi / (2\mathcal{M})\right)^{2/3}$ defines the \emph{KK scale}, with $R_\psi$ being the size of the extra compact dimension involved in forming the smooth topological cap. 

We thus demonstrate that Schwarzschild topological solitons naturally induce strong gravitational effects sharply localized at the horizon scale. These effects trigger sudden tidal disruption and internal excitation of infalling probes, leading to irreversible trapping and efficient scrambling into stringy states. This emerges as a robust feature of these geometries, enabling them to reproduce black hole-like absorption for a broad class of probes with internal degrees of freedom. It provides compelling evidence that an \emph{effective horizon} can arise from a horizonless, smooth microstructure. \\

To simplify the analysis while preserving the essential physics, we work with the neutral, regular bound states of two extremal black holes constructed in \cite{Heidmann:2023kry}, instead of the intricate Schwarzschild topological solitons of \cite{Bah:2020pdz,Bah:2023ows}. These geometries replace the Schwarzschild horizon with a partially smooth, horizonless structure, allowing some probes to interact solely with the cap. As such, the qualitative extension to a fully smooth solution remains straightforward. These configurations thus strike an ideal balance between analytical tractability and physical relevance, enabling a simplified analysis without sacrificing the essential features of the problem.

In Section \ref{section:inital_data}, we review the geometry and introduce coordinates for direct comparison to Schwarzschild. In Sections \ref{sec:Geo} and \ref{sec:TidalStress}, we study the geodesic motion of massive point particles and show that they follow Schwarzschild-like trajectories but encounter extreme tidal forces near the cap. In Section \ref{sec:TidalString}, we extend the analysis to massless strings and show how tidal excitations generate effective mass and dynamically trap the string near $r \sim 2\mathcal{M}$. We conclude in Section \ref{sec:conclusion} with some outlook.
\vspace{-0.3cm}
\subsection*{Summary of results}

For readers primarily interested in the key outcomes, we summarize the main results below. These are also visually represented in Figure \ref{fig:Intro}:
\begin{itemize}\itemsep0em
    \item \textbf{Geodesics:} Free-falling massive particles follow nearly radial Schwarzschild trajectories by slightly deviating after passing the photon sphere at $r \sim 3\mathcal{M}$. Neglecting interaction and internal mode excitation, they bounce off the cap, escaping to infinity (left panel).
    \item \textbf{Tidal Stress:} The tidal forces experienced by these probes diverge from Schwarzschild predictions below $r \sim 3\mathcal{M}$, increasing sharply and peaking at the cap near $r \sim 2\mathcal{M}$. The KK scale and ADM mass set the maximum, such that it reaches extreme string-scale values when $\mathcal{M} > \left( \frac{R_\psi}{l_s}\right)^2 R_\psi$, where $l_s$ is the string length (middle panel). 
    \item \textbf{String probes:} Infalling massless strings experience similar extreme tidal forces that are strong enough to overcome the string tension in a narrow region above the cap (right panel). These tidal forces result in string mode excitations up to a mode cutoff set by the incident energy. As a result, a string that enters as a massless particle emerges in a highly excited, effectively massive state.
    \item \textbf{Trapping Mechanism:} Due to its new effective mass, the string rebounds but fails to escape. Its maximal post-bounce radius lies below $r \sim 2\mathcal{M}(1 + \mathcal{O}(\epsilon,\zeta))$, where $\zeta= \left(\frac{R_\psi^{10}}{\mathcal{M}^4 l_s^6} \right)^{2/3}$ 
    , and, remarkably, is \emph{independent} of the incident energy. The string is therefore confined at the horizon scale, regardless of its initial energy (right panel). This energy-independence is crucial for establishing a universal trapping mechanism in horizonless geometries. It originates from a self-regulating process: as the incident energy increases, additional internal modes are excited during the tidal instability, raising the effective mass and precisely compensating for the added kinetic energy.
    \item \textbf{Infinitesimal-energy escape:} Despite the robustness of the trapping mechanism, strings must exceed a minimum energy to excite internal modes and become trapped. Strings with very low energy, well below the string length, remain unexcited and escape freely. This sets an infrared threshold for effective absorption, which may yield observable consequences.
\end{itemize}

\vspace{-0.4cm}
\section{Neutral bound state of extremal black holes}\label{section:inital_data} 
\vspace{-0.2cm}

The novel solution constructed in \cite{Heidmann:2023kry} describes a neutral and regular bound state of two extremal black holes in minimal six-dimensional supergravity, governed by the action: 
\begin{equation}
    \mathcal{S} \= \frac{1}{16\pi G_6} \int d^6x \sqrt{-\det g} \left(R-\frac{1}{2}|F|^2 \right),\quad \star F=F\,,
\end{equation}
where $F$ is the field strength of a two-form gauge field. The solution is comparable to a four-dimensional Schwarzschild black hole with two extra compact dimensions: it is static, carries no net charges, and is asymptotic to $\mathbb{R}^{1,3} \times$T$^2$. In the interior, the Schwarzschild horizon is replaced by a smooth KK bubble where one of the extra dimensions smoothly degenerates. Two extremal black holes with opposite charges are located at its poles (see Fig.\ref{fig:BS2BMPV.jpg}). These extremal black holes correspond to supersymmetric and anti-supersymmetric Strominger-Vafa black holes \cite{Strominger:1996sh}, whose microscopic origins are well understood in terms of branes (for supersymmetric) and anti-branes (for anti-supersymmetric).

Although the presence of the extremal black holes makes the geometry only partially smooth and horizonless, they are sufficiently localized so that some geodesics avoid them entirely. For such trajectories, the dynamics of massive particles and massless strings effectively probe a fully smooth and horizonless geometry. Moreover, it will be feasible to extend the analysis to geodesics that do interact with the extremal black holes by considering qualitatively more refined fully horizonless geometries where the extremal black holes are further resolved into smooth topological structures.

\begin{figure}[t]
    \begin{center}
    \includegraphics[width=0.35\textwidth]{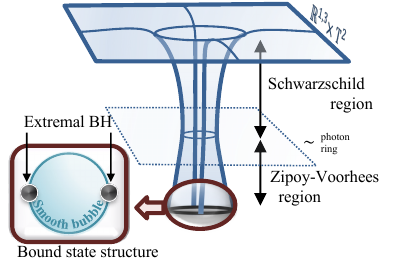}
    \caption{\textit{Spacetime structure of a regular, neutral bound state consisting of two extremal black holes with opposite charges on a smooth Kaluza-Klein bubble.}}
    \label{fig:BS2BMPV.jpg}
    \end{center}
\end{figure}

\subsection{The solution}

The neutral bound state of two extremal black holes is given by the following six-dimensional metric and field strength:
\begin{equation}
\begin{aligned}
d s^2= & -\frac{d t^2}{Z^2}+(d y-T d t)^2  \\
& +Z\left[f d \psi^2+e^{3 \nu}\left(\frac{d \bar{r}^2}{f}+ \bar{r}^2 d \bar{\theta}^2\right)+\bar{r}^2 \sin ^2 \bar{\theta} d \phi^2\right], \\
 F =&\,  dH \wedge d\phi \wedge d\psi -dT\wedge dt \wedge dy \,. \label{eq:6dmetric}
\end{aligned}
\end{equation}
where  $f = 1-\frac{\ell}{\bar{r}}$ and 
\begin{equation}
\begin{aligned}
Z & = 1+\frac{2M(2\bar{r}+M-\ell)}{(2\bar{r}-\ell)^2-\ell^2\cos^2\bar{\theta}-M^2 \sin^2 \bar{\theta}} , \\
T & = \frac{2M \sqrt{\ell^2-M^2}\,\cos\bar{\theta}}{(2\bar{r}+M-\ell)^2-(\ell^2-M^2)\cos^2\bar{\theta}} \,,  \\
 H &=\frac{M \sqrt{\ell^2-M^2}(2\bar{r}+M-\ell)\,\sin^2\bar{\theta}}{(2r-\ell)^2-\ell^2\cos^2\bar{\theta}-M^2 \sin^2 \bar{\theta}}\,, \\
  e^{\nu} & =1-\frac{M^2\sin^2\bar{\theta}}{(2\bar{r}-\ell)^2-\ell^2\cos^2\bar{\theta}} ,  
\end{aligned}
\label{eq:2BHfields} 
\end{equation}
The solution is parameterized by two quantities: $M$, which can be interpreted as the mass contribution from the extremal black holes, and $\ell\geq M$, which sets the bubble size. The coordinates $y$ and $\psi$ parametrize the two compact dimensions, with periods $2\pi R_y$ and $2\pi R_\psi$, respectively.

Although the magnetic and electric gauge potentials, $H$ and $T$, are nontrivially activated, they only contribute as electromagnetic dipoles of order $M\sqrt{\ell^2-M^2}$. The ADM mass, obtained upon dimensional reduction along the T$^2$, is given by (in units where $G_4=1$):
\begin{equation}
\mathcal{M}=\frac{\ell+3 M}{4}. 
\label{eq:Madm}
\end{equation}
The spacetime is regular for $\bar{r}>\ell$ and has a smooth S$^2\times$T$^2$ topology. The surface $\bar{r}=\ell$ corresponds to the ``\emph{cap}" and marks the end of spacetime, where the $\psi$ circle shrinks smoothly as an origin in $\mathbb{R}^2$. The spacetime structure of the solution has been depicted in Fig.\ref{fig:BS2BMPV.jpg}.

Regularity at the cap requires the bubble size $\ell$ to satisfy the following equilibrium condition where the bubble pressure counterbalances the black-hole attraction: 
\begin{equation}
    R_{\psi} = \frac{2(\ell^2-M^2)^\frac{3}{2}}{\ell^2}\,.
\label{eq:RegBu}
\end{equation}
As shown in \cite{Heidmann:2023kry}, the poles of the bubble, $\bar{r}=\ell$ and $\bar{\theta}=0,\pi$, correspond to static extremal black holes with opposite charges, given by $\pm Q=\pm M R_\psi \sqrt{\frac{\ell+M}{\ell-M}}$.\footnote{The charges have a dimension [Length]$^2$ because the black holes are five-dimensional black holes with a S$^3$ horizon topology \cite{Heidmann:2023kry}.} \\

The regularity condition (\ref{eq:RegBu}) imposes a lower bound on the ADM mass and the configuration size, $(2\ell, 8\mathcal{M}) \geq R_\psi$, and we can distinguish two distinct regimes:
\begin{itemize}
    \item \underline{Microscopic bound states:} when $\mathcal{M} \sim \frac{1}{8} R_\psi$, the mass is dominated by the KK bubble, with the extremal black holes being negligible, $M\ll R_\psi$. We call such a solution ``\emph{microscopic}'' as the ADM mass and size of the bound state scale with the radius of the compact dimension, $\psi$.
    \item \underline{Macroscopic bound states:} when $\mathcal{M} \gg R_\psi$, the mass is primarily sourced by the extremal black holes. At leading order in $\frac{R_\psi}{\mathcal{M}} \ll 1$, we find
    \begin{equation}
    \ell=\mathcal{M}\left(1+\frac{3 \epsilon}{8}\right), \quad M =\mathcal{M}\left(1-\frac{\epsilon}{8}\right),
    \label{Eq:l&M@EPSorder}
\end{equation}
where we have introduced the \emph{Kaluza-Klein scale}, 
\begin{equation}
    \epsilon \equiv\left(\frac{R_\psi}{2 \mathcal{M}}\right)^\frac{2}{3} \ll 1. \label{eq:Defepsilon}
\end{equation}
We refer to such a solution as ``\emph{macroscopic}'' because the ADM mass and size of the bound state decouple from the radius of the compact dimension.
\end{itemize}

\subsection{Zipoy–Voorhees geometry as an approximation above the cap}

For macroscopic geometries and slightly away from the cap, the geometry is indistinguishable from the vacuum $\delta=2$ Zipoy-Voorhees (ZV) solution \cite{zipoy1966topology,*voorhees1970static,*stephani2009exact,Kodama:2003ch}\footnote{It is also referred to as the ``$q$-metric'' in some works \cite{Toktarbay:2014yru}.} embedded trivially in six dimensions (see Fig.\ref{fig:BS2BMPV.jpg}). Indeed, for $\bar{r}\gtrsim \mathcal{M}(1+\mathcal{O}(\epsilon^\frac{1}{4}))$, the electromagnetic fields vanish at leading order in $\epsilon$, and the metric simplifies to 
$ds_6^2 \sim ds_\text{ZV}^2 + d\psi^2 + dy^2$,
where $ds_\text{ZV}^2$ is the ZV metric:
\begin{equation}
\begin{aligned}
ds_\text{ZV}^2 =& -f_{\text{ZV}}^2 \,dt^2+ \frac{\bar{r}^2\sin^2 \bar{\theta}}{f_{\text{ZV}}}\,d\phi^2 \label{eq:DisSchw} \\
&+\frac{f_{\text{ZV}}^2}{\left(f_{\text{ZV}}+\frac{\mathcal{M}^2 \sin^2\bar{\theta}}{4\bar{r}^2}\right)^{3}} \left(\dfrac{d\bar{r}^2}{f_{\text{ZV}}}+\bar{r}^2 d\bar{\theta}^2 \right),
\end{aligned}
\end{equation}
where $f_{\text{ZV}} = 1-\frac{\mathcal{M}}{\bar{r}}$.

The ZV solution can be regarded as an axial S$^2$ deformation of the Schwarzschild metric, inducing a curvature singularity at the horizon, $\bar{r}=\mathcal{M}$. As discussed in \cite{Heidmann:2023kry}, this solution resembles a Schwarzschild black hole up to the photon sphere, below which the deformation becomes manifest (see Fig.\ref{fig:BS2BMPV.jpg}). The bound state reviewed here resolves the singular ZV horizon into a smooth KK bubble with two extremal black holes at its poles.

Given that the ZV metric provides an excellent approximation to the macroscopic bound states of extremal black holes from a KK scale above the cap up to infinity — and given its relative simplicity — we will also analyze probe dynamics within this singular background. This analysis captures the behavior of probes in the corresponding bound state geometry before they reach the cap region.

\subsection{ACMC coordinates}

In this paper, we compare the bound state of extremal black holes to the Schwarzschild black hole. To do so meaningfully, we require a coordinate system that allows for consistent comparison between the solutions. For  Schwarzschild, the $(r, \theta, \phi, t)$ coordinate system serves both as the standard spherical coordinate system adapted to the spherical symmetry where $r$ measures the radius of the sphere and $(\theta, \phi)$ the angular position, and as the ``asymptotically Cartesian and mass-centered'' (ACMC) coordinates \cite{Thorne:1980ru}.

In contrast, the coordinates $(\bar{r}, \bar{\theta})$ do not represent the same physical quantities. In the absence of spherical symmetry, we introduce new ACMC coordinates $(r,\theta)$ for both the bound state of extremal black holes and the ZV spacetime, such that any point specified by $r = \text{cst}$, $\theta = \text{cst}$, and $\phi = \text{cst}$ can be meaningfully compared to the same location in the Schwarzschild geometry.

\subsubsection{ACMC coordinates of the ZV geometry}

\begin{figure}[t]
    \begin{center}
    \includegraphics[width=0.48\textwidth]{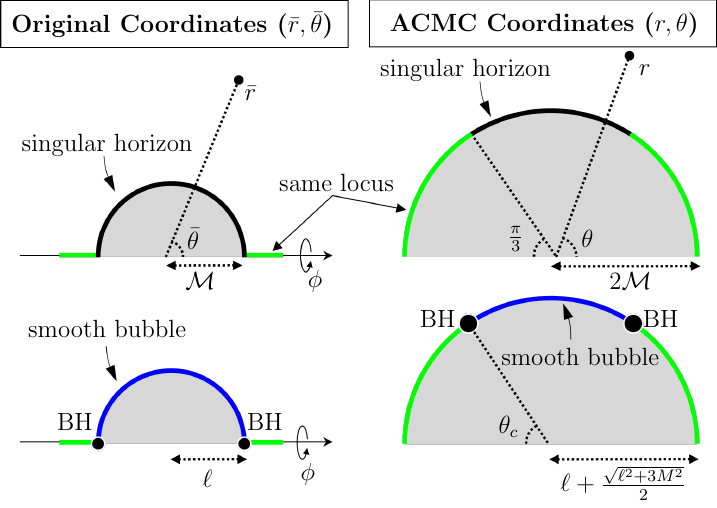}
    \caption{\textit{Correspondence between the  original coordinates $(\bar{r}, \bar{\theta})$ and ACMC coordinates $(r, \theta)$  for both the ZV geometry $($top row$)$ and the bound state $($bottom row$)$.}}
    \label{fig:Original/ACMCcorresp.jpg}
    \end{center}
\end{figure}

ACMC coordinates are obtained by imposing certain constraints on the asymptotic behavior of the metric components (see \cite{Bena:2020uup} for a short review). One such condition is the absence of terms of the form $P_n(\cos \theta)/r^n$ in the asymptotic expansion of $g_{rr}$ at all orders. As a result, the ACMC coordinates typically take a complicated form as a series expansion in the original $(\bar{r}, \bar{\theta})$ coordinates. To keep the expressions manageable while enabling meaningful comparison with the Schwarzschild coordinate system, we restrict ourselves to ACMC-4 coordinates that satisfy the ACMC conditions up to fourth order. For the ZV metric \eqref{eq:DisSchw}, the ACMC-4 coordinates are derived following \cite{Bena:2020uup,Thorne:1980ru} and are related to the original $(\bar{r},\bar{\theta})$ coordinates as follows:}
\begin{equation}
\begin{split}
    &\sqrt{\bar{r}(\bar{r}-\mathcal{M})} \sin \bar{\theta} = \sqrt{r(r-2\mathcal{M})}\sin{\theta} \,,\\
    &\left(\bar{r}-\frac{\mathcal{M}}{2} \right) \cos \bar{\theta} = \left(r-\mathcal{M}\right) \cos{\theta}.
\end{split}
\label{eq:ACMCZV}
\end{equation}
We have illustrated the ZV solution in both the original and ACMC coordinate patches in the top row of Fig.\ref{fig:Original/ACMCcorresp.jpg}.

In the original coordinates, the ZV spacetime is defined for $\bar{r} \geq \mathcal{M}$, with a singular horizon located at $\bar{r} = \mathcal{M}$ and no spacetime below this locus (gray region in Fig.\ref{fig:Original/ACMCcorresp.jpg}). By contrast, the ACMC coordinates are defined for $r \geq 2\mathcal{M}$.

The horizon now lies at $r = 2\mathcal{M}$ (black line in Fig.\ref{fig:Original/ACMCcorresp.jpg}), but it only covers the angular range $\theta \in [\frac{\pi}{3}, \frac{2\pi}{3}]$. The remaining part of the $r = 2\mathcal{M}$ surface (shown in green) corresponds to portions of the symmetry axis above the horizon, where the geometry is smooth and corresponds to the range $\mathcal{M} < \bar{r} \leq \frac{3\mathcal{M}}{2}$ with $\bar{\theta} = 0$ or $\pi$.

Thus, the symmetry axis, where the $\phi$ circle smoothly degenerates, is not simply described by $\theta = 0,\pi$ in the ACMC coordinates. It splits into multiple regions: $\theta = 0,\pi$ for $r > 2\mathcal{M}$ and $\theta \in [0, \frac{\pi}{3}] \cup [\frac{2\pi}{3}, \pi]$ for $r = 2\mathcal{M}$.

As shown in Fig.\ref{fig:Original/ACMCcorresp.jpg}, the ACMC coordinates of the ZV metric appropriately resize the geometry such that the horizon is located at $r = 2\mathcal{M}$. However, the geometry is effectively compressed toward the equatorial plane, such that the ``poles'' of the horizon are now located at $\theta = \pi/3$ and $\theta = 2\pi/3$.

\subsubsection{ACMC coordinates of the bound state}

Applying the same procedure, we find that the ACMC-4 coordinates of the bound state of extremal black holes \eqref{eq:6dmetric} are related to the original coordinates by
\begin{equation}
\begin{split}
\sqrt{\bar{r}(\bar{r}-\ell)} \sin \bar{\theta} &\= \sqrt{\left(r-r_+\right)\left(r-r_-\right)}\sin{\theta}  \\
\left(\bar{r}-\frac{\ell}{2} \right) \cos \bar{\theta} &\= \left(r-\frac{r_++r_-}{2} \right) \cos{\theta},
 \end{split}
\end{equation}
where 
\begin{equation}
    r_\pm \equi \ell \pm \frac{\sqrt{\ell^2+3M^2}}{2}\,.\label{eq:Defrpm}
\end{equation}
The cap, $\bar{r} = \ell$, now corresponds to $r = r_+$ and does not directly depend on the ADM mass. This dependence is expected: unlike the ZV solution, the size of the bound state is determined by its internal topological structure and not solely by its ADM mass \eqref{eq:Madm}. One can show that
\begin{equation}
    2\mathcal{M} \,<\, r_+ \,\leq\, 4\mathcal{M}.
\end{equation}
More precisely, the bound state is twice less compact than the Schwarzschild black hole in the microscopic limit $\mathcal{M} \sim \frac{1}{8} R_\psi$, and asymptotically approaches the Schwarzschild size in the macroscopic limit \eqref{Eq:l&M@EPSorder},
\begin{equation}
    r_+ \= 2\mathcal{M}\left(1+\frac{3\epsilon}{16} \right).
\end{equation}

Therefore, in the macroscopic limit, the regime of interest, the ACMC coordinates reduce to those of \eqref{eq:ACMCZV} and the bound state is as compact as the Schwarzschild black hole, up to small corrections set by the KK scale.

We have depicted the bound-state spacetime in both the original and ACMC coordinate patches in the bottom row of Fig.\ref{fig:Original/ACMCcorresp.jpg}.

As with the ZV solution, the $r = r_+$ locus splits into three regions. The central region $\theta \in [\theta_c, \pi - \theta_c]$ corresponds to the bubble locus at $\bar{r} = \ell$, with the two extremal black holes located at its extremities, where 
\begin{equation}
    \cos \theta_c \= \frac{\ell}{\sqrt{\ell^2+3M^2}}.
\end{equation}
As expected, $\theta_c \to \pi/3$ in the macroscopic limit \eqref{Eq:l&M@EPSorder}. The complementary region $\theta \in [0, \theta_c] \cup [\pi - \theta_c, \pi]$ corresponds to the symmetry axis $\bar{\theta} = 0, \pi$ in the radial range $\ell < \bar{r} \leq \frac{1}{2}(\ell + \sqrt{\ell^2 + 3M^2})$.

Thus, as with the ZV solution, the ACMC coordinates of the bound state yield a coordinate system comparable to that of Schwarzschild, while inducing a nontrivial distortion in the cap region. The spacetime ends at $r = r_+ \gtrsim 2\mathcal{M}$, where either the $\phi$ circle or the $\psi$ circle smoothly degenerates, corresponding respectively to the symmetry axis and the bubble locus. The two extremal black holes, located at the junctions, are no longer positioned at the poles of the ACMC coordinate system but are instead ``pushed" toward the equator, as illustrated in Fig.\ref{fig:Original/ACMCcorresp.jpg}.

\section{Free fall of massive particles}\label{sec:Geo}

In this section, we derive kinematic properties of massive point particles in free fall within the backgrounds reviewed in the previous section and compare them to infall in the Schwarzschild black hole. Massive probe particles of mass $m$ follow geodesic trajectories governed by the Hamiltonian:
\begin{equation}
    H=\frac{1}{2} g^{\mu \nu} \,p_\mu p_\nu=-\frac{m^2}{2}, \qquad p_\mu\equiv g_{\mu \nu} \,\dot{x}^\nu
    \label{eq:constOfMotion}
\end{equation}
where $p_\mu$ are the conjugate momenta, and $\dot{x}^\nu = \frac{dx^\nu}{d\tau}$ are the coordinate derivatives of the probe with respect to the affine parameter $\tau$. The geodesic equations are obtained from the Hamilton-Jacobi equations:
\begin{equation}
    \dot{p}_\mu=-\frac{d H}{d x^\mu}.
    \label{eq:HamilEqs}
\end{equation} 
Since the geometries exhibit four isometries along the time, azimuthal, and compact dimensions, the momenta along these directions are conserved, implying $\dot{p}_t=\dot{p}_\phi=\dot{p}_y=\dot{p}_\psi=0$. 

We consider freely falling particles with $p_\phi=0$. Additionally, we restrict ourselves to the lowest Kaluza-Klein modes by setting $p_y=p_\psi=0$.\footnote{As argued in \cite{Heidmann:2022ehn}, momenta along extra dimensions behave as an effective large mass term from a four-dimensional perspective due to their quantization in $R_{y/\psi}^{-1}$.} Finally, we choose the affine parameter $\tau$ to be the proper time, so that $p_t=-E$, where $E$ is the energy of the probe. Under these conditions, the equations governing the infall motion of massive or massless particles in our six-dimensional backgrounds are: 
\begin{equation}
\begin{split}
    &\dot{\phi} = \dot{\psi} = 0 ,\qquad \dot{t}=-g^{tt} E,\qquad \dot{y}=-g^{ty} E,\\
    &2H \= g_{rr}\,\dot{r}^2+g_{\theta\theta}\,\dot{\theta}^2+g^{tt} E^2\= -m^2\,.
\end{split}
    \label{eq:GeoEqGen}
\end{equation}
where $m=1$ (resp. $0$) for timelike (resp. null) geodesics.

Since we are interested in massive probes released at rest from a large distance, $r_0$, we have $E^{-2}=-g^{tt}|_{r=r_0}\sim 1$, and we can consider $E=1$ for simplicity.

For the Schwarzschild black hole trivially embedded in six dimensions, spherical symmetry imposes a purely radial free-fall trajectory, meaning $\dot{\theta}=0$. The geodesic motion then follows the well-known equations:
\begin{equation}
    \dot{r} \= -\sqrt{\frac{2\mathcal{M}}{r}}\,,\qquad \dot{t} \= \left(1-\frac{2\mathcal{M}}{r}\right)^{-1},
    \label{eq:GeoSchw}
\end{equation}
and the proper time, $\Delta \tau_\text{Schw}$, it takes to a particle starting at $r_0 \gg 2\mathcal{M}$ to reach the horizon is
\begin{equation}
    \Delta \tau_\text{Schw} \= \frac{4\mathcal{M}}{3} \left[ \left( \frac{r_0}{2\mathcal{M}} \right)^\frac{3}{2} -1 \right]. \label{eq:SchwarzschildGeo}
\end{equation}

Due to the axial symmetry of our solutions, free-falling particles follow more intricate trajectories, which we will derive and analyze mainly numerically. However, since the solutions exhibit a $\mathbb{Z}_2$ symmetry under $\theta \to \pi -\theta$, the geodesics on the equatorial plane at $\theta = \pi/2$ remain on the equatorial plane. Thus, analytic derivation will be provided for probes starting at $\theta=\pi/2$, four which the four-dimensional trajectory is governed by: $$ \dot{\theta} = 0 ,\qquad \dot{t}= - g^{tt}|_{\theta=\frac{\pi}{2}},\qquad \dot{r}=\sqrt{\frac{-g^{tt}-1}{g_{rr}}}\Bigl|_{\theta=\frac{\pi}{2}}.$$

\subsection{In the Zipoy-Voorhees geometry}

We first consider infalling massive probes in the ZV spacetime \eqref{eq:DisSchw}, which closely resembles the bound state of extremal black holes in the macroscopic regime $\mathcal{M} \gg R_\psi$. On the equatorial plane, $\theta=\pi/2$, the geodesic equations yield:
\begin{equation}
\begin{aligned}
    \dot{r}^2 &= \frac{\mathcal{M}h_{\text{ZV}}^9\left(h_{\text{ZV}}-\mathcal{M}\right)^2}{2^{10} r^5(r-\mathcal{M})^2(r-2\mathcal{M})^5},\quad
    \dot{t} = \frac{\left(h_{\mathrm{ZV}}+\mathcal{M}\right)^2}{\left(h_{\mathrm{ZV}}-\mathcal{M}\right)^2},
    \label{eq:GeoZV}
\end{aligned}
\end{equation}
where $h_{\text{ZV}} \equiv \sqrt{4 r(r-2\mathcal{M})+\mathcal{M}^2}$.

To compare with the Schwarzschild free-fall, we derive the proper time, $\Delta \tau_\text{ZV}$, required for the probe to reach the horizon at $r=2\mathcal{M}$ starting from $r_0 \gg 2\mathcal{M}$:
\\
\begin{equation}
\Delta \tau_{\text{ZV}}= \Delta \tau_{\text {Schw}} - 0.26\mathcal{M},
\label{Eq:ZVtime}
\end{equation}
where we have approximated the finite coefficient that has an explicit expression in terms of elliptic integrals.

Thus, the infall time differs from the Schwarzschild value by a relatively small amount, reflecting the fact that the ZV solution closely resembles Schwarzschild spacetime up to the photon orbit at $r \sim 3\mathcal{M}$, with deviations appearing only beyond this region. For a qualitative comparison, the proper time for a massive particle to reach the surface of a regular star located at $r = R_S$ agrees with \eqref{Eq:ZVtime} when $R_S = 1.13 \times 2\mathcal{M}$.\\

\begin{figure*}[t]
    \begin{center}
    \includegraphics[width=0.95\linewidth]{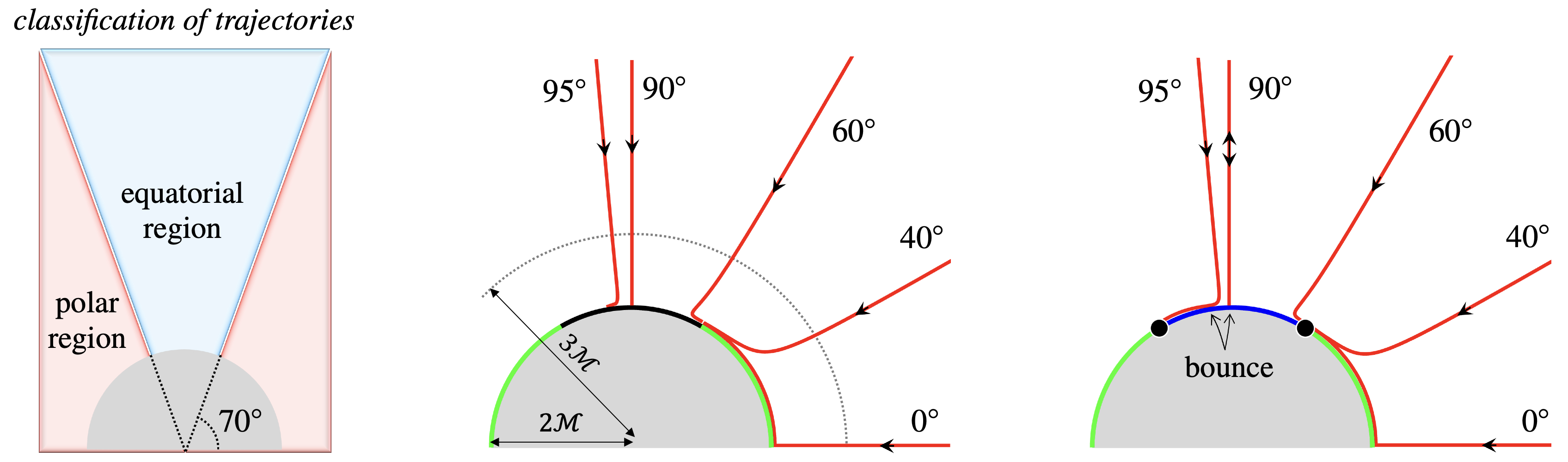}
    \caption{\textit{Kinematic properties of freely falling massive probes in the ZV spacetime and a neutral bound state of extremal black holes with $\epsilon=0.2$. Left panel: Classification of release angles according to whether the probe reaches the poles of the geometry first (polar region) or falls closer to the equatorial locus (equatorial region). Right panels: Representative trajectories in both geometries, illustrating the qualitative similarity of the motion and the key differences near the smooth bubble.}}
    \label{fig:ProbeRegions}
    \end{center}
\end{figure*}

We now consider generic infall trajectories with arbitrary release angles. The motion is not purely radial, and $\theta$ evolves. The geodesic equations governing $(r(\tau),\theta(\tau))$ follow from \eqref{eq:HamilEqs} with the Hamiltonian \eqref{eq:GeoEqGen}. We solve these equations numerically to determine the trajectories $(r(\tau),\theta(\tau))$ for various release angles, $0\leq \theta_0 < \pi$, and for an initial radial position $r_0 \gg 2 \mathcal{M}$. Figure \ref{fig:ProbeRegions} presents the key results, along with representative trajectories.

First, the middle panel clearly shows that, despite the axial symmetry, trajectories remain nearly radial down to $r \sim 3\mathcal{M}$, with noticeable deviations only emerging below this radius. As in the case of geodesics on the equatorial plane, the infall of massive particles closely mirrors that in Schwarzschild spacetime, with only minor deviations appearing beneath the photon orbit.

Second, all trajectories end at the singular horizon. However, the poles of the horizon act as strong attractors. Specifically, we find that particles released within the angular range $\theta_0 \in [0, 70^\circ] \cup [110^\circ, 180^\circ]$, which we refer to as the \emph{polar region}, ultimately fall into one of the poles of the ZV horizon (left panel in Fig.\ref{fig:ProbeRegions}). Consequently, only particles within the  \emph{equatorial region}, $70^\circ < \theta_0 < 110^\circ$, reach the $r=2\mathcal{M}$ locus between the poles.

This strong polar attraction is consistent with expectations: the ZV metric approximates a bound state of two extremal black holes supported by a KK bubble, and in such a configuration, massive particles are naturally drawn more strongly to the black holes than to the bubble. What is remarkable, however, is that this feature survives in the ZV geometry, even though the bound-state structure has been completely smeared out.

In summary, the basic kinematic behavior of freely falling massive particles in the ZV spacetime is nearly indistinguishable from that in Schwarzschild, with mild differences emerging only between the Schwarzschild photon orbit and the horizon.

\subsection{In the bound state of extremal black holes}
\label{sec:TrajBS}

We now turn to infalling massive probes in the bound-state geometry \eqref{eq:6dmetric}. On the equatorial plane, radial infall is governed by the geodesic equations:\footnote{We omit the motion along the $y$ direction, $\dot{y}=-g^{ty}|_{\frac{\pi}{2}}$, as for macroscopic solutions \eqref{Eq:l&M@EPSorder}, one has $\dot{y} = \mathcal{O}(\sqrt{\epsilon})$, and therefore negligible.}
\begin{equation}
   \medmath{ \dot{r}^2 = \frac{M h_{\text{BS}}^9 (h_{\text{BS}}-\ell)}{2^8 (\ell+h_{\text{BS}})(r-\ell)^2((r-\ell)^2-M^2)^4} ,\quad \dot{t} = \frac{(h_{\text{BS}}+M)^2}{(h_{\text{BS}}-M)^2},}
    \label{eq:GeoBS}
\end{equation}
where $h_{\text{BS}} \equiv \sqrt{4(r-\ell)^2-3M^2}$. 

The proper time $\Delta \tau_\text{BS}$ required for a probe to reach the cap at $r = r_+$ from an initial position $r_0 \gg r_+$ is
\begin{equation}
\begin{split}
  \Delta \tau_{\text{BS}}&\= \frac{4M}{3} \left(\left(\frac{r_0}{2M} \right)^\frac{3}{2} - C_{M,\ell} \right),\\
  C_{M,\ell}&\equi \medmath{\frac{6M^2(5\ell^2-M^2) \,E(\tfrac{1}{2})+(5\ell^4-M^4)\left(3E(\tfrac{1}{2})-2 K(\tfrac{1}{2})\right)}{20\sqrt{2\ell^5 M^3}},}
\end{split}
\end{equation}
where $K$ and $E$ are the complete elliptic integrals of the first and second kind, respectively, with $E(\tfrac{1}{2})\approx 1.35$ and $K(\tfrac{1}{2})\approx 1.85$.

Unlike the Schwarzschild or ZV solutions, the infall time in this background depends primarily on $M$, the mass contribution of the extremal black holes, rather than on $\mathcal{M}$, the ADM mass of the full geometry. This behavior is characteristic of spacetimes with topological structure from extra-dimensional deformations: the large-distance gravitational potential, governed by $-g^{tt} - 1$, does not necessarily coincide with the ADM mass extracted from the four-dimensional reduction, $-g^{(4d)}_{tt} - 1$.

The parameter $\ell$ encodes the topological size of the Euclidean Schwarzschild bubble and does not contribute to the gravitational potential. In the limit $M \to 0$, the background reduces to a six-dimensional Euclidean Schwarzschild geometry with no redshift ($g^{tt} = -1$), implying $\dot{r} = 0$ and no gravitational attraction. Hence, these microscopic configurations exhibit the exotic behavior of topological geometries, sharply contrasting with Schwarzschild black holes.

In contrast, macroscopic bound states defined by \eqref{Eq:l&M@EPSorder}, with $M \sim \mathcal{M}$, begin to exhibit Schwarzschild-like behavior. To leading order in $\epsilon$, the proper time becomes 
\begin{equation}
    \Delta \tau_{\text{BS}} \= \Delta \tau_{\text{ZV}} + \epsilon \mathcal{M} \left(\left(\frac{r_0}{2\mathcal{M}}\right)^\frac{3}{2} + 0.12 \right).
\end{equation}
Although the parameter driving gravitational attraction, $M$, induces corrections proportional to $r_0/(2\mathcal{M})$, it is suppressed by $\epsilon$, ensuring that the kinematics remain nearly indistinguishable from those in the ZV or Schwarzschild geometries.

However, a drastic difference emerges near $r = r_+$, at the smooth Kaluza-Klein bubble. At this locus, the $\psi$ circle pinches off like the ``tip of a cigar," and infalling probes are smoothly reflected by undergoing a transition $\psi \to \psi + \pi R_\psi$, effectively \emph{bouncing off} the bubble and escaping back to infinity. This behavior is radically different from infall into a horizon and is characteristic of ultra-compact, horizonless geometries where the horizon is replaced by smooth topological structure. For our bound states, this bubble is smooth almost everywhere except at its poles (see Fig.~\ref{fig:Original/ACMCcorresp.jpg}), so such bouncing behavior is expected. \\

We now turn to generic free-fall trajectories with arbitrary release angles away from the equatorial plane. As in the ZV geometry, the motion acquires a non-radial component, and the angle $\theta$ evolves. We integrate the geodesic equations \eqref{eq:HamilEqs} numerically to obtain the trajectories $(r(\tau), \theta(\tau))$ for a range of release angles $0\leq \theta_0<\pi$, assuming an initial radial position $r_0 \gg 2 \mathcal{M}$. The results are shown in the right panel of Fig.\ref{fig:ProbeRegions} for an illustrative macroscopic bound state with $\epsilon =0.2 $ .

As expected, all trajectories closely follow those of the ZV geometry, with noticeable differences arising only near the cap, where the spacetimes start to differ. Particles released in the \emph{polar region}, as defined in the ZV background, similarly reach the poles of the bubble first, where the black holes are, and are absorbed there. In contrast, those in the \emph{equatorial region} fall toward the smooth region of the bubble. Unlike in the ZV geometry, these trajectories undergo a bounce at the smooth surface, just as in the equatorial plane.

However, away from the equatorial plane, the bounce is not symmetric: the probe becomes increasingly attracted to one of the black holes. As a result, after bouncing on the bubble, the particle is typically pulled toward and ultimately absorbed by one of the black holes (see Fig.\ref{fig:ProbeRegions}).

Thus, the black holes at the poles act as strong gravitational sinks. All massive probes in the polar region fall directly into one of them, while those originating from the equatorial region also end up in the black holes after a single bounce, except for the special case of probes released exactly in the equatorial plane, which escape the geometry. This occurs because the Schwarzschild horizon has been only partially replaced by a smooth topological structure. The geometry still contains localized black holes, which, despite their confinement to small regions, dominate the gravitational potential.

If we had instead considered the more intricate, fully smooth horizonless geometries constructed in \cite{Bah:2022yji,Bah:2023ows}, where the black holes are replaced by small charged KK bubbles, all probes would bounce off the smooth surfaces and escape the geometry due to energy conservation. While the analysis of these geometries would be more involved, it would not alter this essential conclusion.

In summary, the basic kinematic properties of massive particles in free fall within our bound-state background, and by extension to the Schwarzschild topological solitons of \cite{Bah:2022yji,Bah:2023ows}, closely resemble those in the ZV and Schwarzschild geometries. The critical distinction lies in the fate of particles near $r=r_+ \sim 2\mathcal{M}$: the replacement of the horizon by a smooth topological bubble introduces the possibility of a bounce and escape. This constitutes a significant deviation from standard black hole dynamics. As we will see, the inclusion of tidal stresses will further refine this picture, restoring an effective trapping mechanism even in the absence of a true event horizon.

\section{Tidal Stresses}
\label{sec:TidalStress}

We now turn to a more detailed analysis of the infall dynamics, focusing on tidal forces and geodesic deviation. Although it is commonly said that there is ``no drama at the horizon of a black hole,'' it has been shown for supersymmetric horizonless geometries, that spacetimes ending as a smooth highly-redshifted cap can significantly alter this picture and induce extreme tidal stresses on probes way before they reach the cap \cite{Tyukov:2017uig,Bena:2018mpb,Bena:2020iyw}.

We perform a similar analysis for our astrophysically relevant Schwarzschild-like geometries. We show that while these solutions retain key features of their supersymmetric counterparts, such as the amplification of tidal stresses, the extreme values capable of triggering new physics are only reached near the cap, corresponding to the horizon scale of the geometries, $r\sim 2\mathcal{M}$.

The tidal forces experienced by an infalling massive particle are determined by the geodesic deviation equation along its trajectory: 
\begin{equation}
    \frac{D^2 S^\mu}{d\tau^2} = 
    \mathcal{A}^\mu{}_\nu S^\nu,\qquad \mathcal{A}^\mu{}_\nu \equiv - R^\mu{}_{\rho \nu\sigma}v^\rho v^\sigma,
    \label{eq:EqOfGeo}
\end{equation}
where $D$ stands for the covariant derivative along the curve parameterized by proper time,  $v^\mu=\dot{x}^\mu$ is the proper velocity of the particle, and $R^\mu{}_{\rho \nu \sigma}$ is the Riemann curvature tensor. For our six-dimensional geometries with geodesic equations for free-falling particles given by \eqref{eq:GeoEqGen}, the proper velocity gives
\begin{equation}
    v^\mu \= \left(\dot{t},\dot{r},\dot{\theta},\dot{\phi},\dot{\psi},\dot{y}\right) \= \left(-g^{tt},\dot{r},\dot{\theta},0,0,-g^{ty}\right)\,,
\end{equation}
where $(\dot{r},\dot{\theta})$ are governed by geodesic equations \eqref{eq:HamilEqs}.

The vector $S^\mu$ represents the infinitesimal separation between two neighboring geodesics. In the rest frame of the reference geodesic, $S^\mu$ is spacelike, and describes how nearby particles ``deviate'' due to tidal effects. As such, the tensor $\mathcal{A}$ is referred to as the \emph{tidal tensor}.

\begin{figure*}[t]
    \begin{center}
    \includegraphics[width=0.94\linewidth]{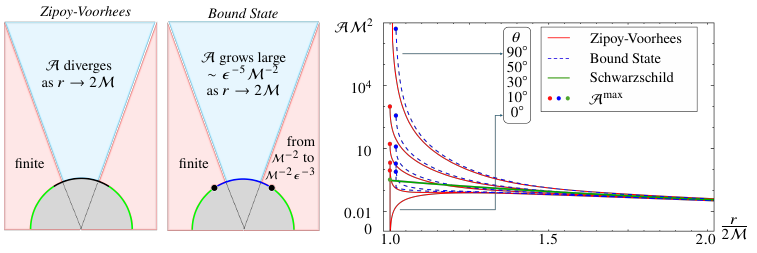}
    \caption{\textit{Characteristics of the tidal forces exerted on free-falling massive probes in the ZV and bound-state spacetimes. Left panels: Behavior of the maximal tidal force for probes falling from the equatorial and polar regions. Right panel: Tidal forces as a function of the radial distance for Schwarzschild, ZV, and a bound state with $\epsilon= 0.1$, and for various release angles.}}
    \label{fig:TidForcesComb}
    \end{center}
\end{figure*}

To quantify the overall strength of the tidal forces along the trajectory, we introduce the norm of the tidal tensor:
\begin{equation}
    \mathcal{A} = \sqrt{\mathcal{A}^\mu{}_\nu \mathcal{A}^\nu{}_\mu}.
\end{equation}

For a radial free-fall trajectory in the Schwarzschild background, we retrieve the known result
\begin{equation}
    \mathcal{A}_\text{Schw} = \frac{\sqrt{6}\mathcal{M}}{r^3}
    \label{eq:ASchw}
\end{equation}
showing that tidal forces remain mild outside the horizon when the black hole mass is large. In fact, the maximum tidal stress is given by $\mathcal{A}_\text{Schw}^\text{max} = \frac{\sqrt{6}}{8\mathcal{M}^2}$. \\

As for the probe trajectories, the axial symmetry of the ZV spacetime or the bound state makes analytical derivation possible only on the equatorial plane, $\theta=\pi/2$, where we have:
\begin{equation}
    v^\mu \= \left(-g^{tt},\sqrt{\frac{-g^{tt}-1}{g_{rr}}},0,0,0,-g^{ty}\right)\Bigg|_{\theta=\frac{\pi}{2}}.
\end{equation}
In the following, we compute analytically the tidal forces experienced by massive particles on the equatorial plane and provide a numerical derivation outside the plane.

\subsection{In the Zipoy-Voorhees geometry}

On the equatorial plane of the ZV spacetime, the norm of the tidal tensor for a free-falling massive particle takes the form, as a function of $r$,
\begin{equation}
    \mathcal{A}_\text{ZV}|_{\frac{\pi}{2}} \= \mathcal{A}_\text{Schw} \,\,\frac{g\left(\frac{r}{2\mathcal{M}}\right)}{\left(1-\frac{2 \mathcal{M}}{r}\right)^5}\,,
\end{equation}
where $g$ is a nontrivial, monotonic function of $\frac{r}{2\mathcal{M}}$ that starts at $\sqrt{3}/16384$ when evaluated at $1$ and asymptotically approaches $1$ at large distances.

The tidal force diverges at $r = 2\mathcal{M}$. Although the leading divergence follows $(r - 2\mathcal{M})^{-5}$, a closer examination of the multiplicative factor $g$ reveals that deviations from the Schwarzschild tidal force only become significant below the photon orbit, $r< 3\mathcal{M}$. Specifically, we find $\mathcal{A}_\text{ZV}|_{\frac{\pi}{2}} \sim 1.1 \, \mathcal{A}_\text{Schw}$ at $4\mathcal{M}$ and $\mathcal{A}_\text{ZV}|_{\frac{\pi}{2}} \sim 1.45 \, \mathcal{A}_\text{Schw}$ at $3\mathcal{M}$.  \\

To analyze the tidal stress outside the equatorial plane, we numerically solve the geodesic equations, from which we compute the corresponding velocities and subsequently the tidal tensor. The results are shown in Fig.\ref{fig:TidForcesComb}.

The right plot demonstrates that the tidal stress begins to deviate significantly from the Schwarzschild value beyond the photon orbit and eventually reaches much higher magnitudes. The most extreme stress occurs on the equatorial plane, while the minimum is observed for probes falling along the symmetry axis, $\theta = 0,\pi$.

More precisely, we find that the tidal force diverges at the horizon only for probes released from the equatorial region, which reaches the singular locus in between its poles. In contrast, probes descending from the polar regions experience large but finite tidal stresses.

A particularly interesting case is the geodesic falling along the symmetry axis, $\theta = 0$ or $\pi$, whose trajectory is shown in Fig.\ref{fig:ProbeRegions}. The tidal force vanishes when reaching the $r = 2\mathcal{M}$ locus and then grows as the probe follows the green locus up to the poles of ZV horizon. This vanishing of the tidal force is a remarkable and unusual feature. Although the spacetime remains curved, the structure of the Riemann tensor and the direction of geodesic motion conspire to cancel tidal acceleration. As a result, the probe temporarily experiences no differential gravitational stretching or compression, despite being immersed in a varying gravitational field.

In summary, while the kinematic properties of free-falling probes in the ZV spacetime deviate only mildly from those in the Schwarzschild black hole, the situation is markedly different when considering tidal stresses. We have shown that the tidal force can become extremely large between the horizon and the photon orbit and even diverge along a significant set of infalling trajectories. While the large variation in maximal tidal forces depending on the release angle may seem puzzling, we will clarify this behavior in the next section by relating it to the presence of the two extremal black holes.

\subsection{In the bound state of extremal black holes}
\label{sec:TidalParticleBS}

For equatorial probes, the norm of the tidal force as a function of radial distance can be analytically computed for the bound state of extremal black holes. In the case of microscopic bound states ($\mathcal{M} \sim \tfrac{1}{8} R_\psi$), the tidal force is negligible throughout, consistent with the fact that such geometries exert very little gravitational attraction. However, for macroscopic bound states satisfying \eqref{Eq:l&M@EPSorder}, the tidal force mirrors that of the ZV spacetime, but, in the absence of a singularity, the divergence is resolved. The tidal force reaches a large, finite value at the bubble locus $r = r_+$, given to leading order in $\epsilon$ as: 
\begin{equation}
    \mathcal{A}^\text{max}|_{\theta = \frac{\pi}{2}}\= 19^{\frac{1}{2}}\left(\frac{2^{19} \mathcal{M}^{4}}{R_\psi^{10}}\right)^\frac{1}{3} \=  \frac{8\sqrt{19}}{\epsilon^5 \mathcal{M}^2} .
\label{eq:EquatStressBS}
\end{equation}
Consequently, the tidal stress attains extreme values in the vicinity of $r\sim 2\mathcal{M}$, determined by the KK scale and the ADM mass.

Moreover, a consistent supergravity description requires both $R_\psi$ and $\mathcal{M}$ to be much larger than the six-dimensional Planck length $l_p$ and the string length $l_s = \sqrt{\alpha'}$. Anticipating our analysis of infalling massless strings, we will express all relevant quantities in units of the string length. The tidal stress discussed above reaches the string scale, $\mathcal{A}^\text{max}|_{\theta = \frac{\pi}{2}} > \alpha'^{-1}$, whenever $\epsilon^5 \mathcal{M}^2 < \alpha'$. This condition is met when the ADM mass satisfies the following nonrestrictive bound:
\begin{equation}
    \left(\frac{R_\psi}{l_s} \right)^{3/2} R_\psi \,<\, \mathcal{M}, \label{eq:ADMMassBoundfirst}
\end{equation}
To ensure that the tidal stress reaches \emph{extreme} string-scale values, we impose a stronger condition:
\begin{equation}
    \left(\frac{R_\psi}{l_s} \right)^2\, R_\psi \,<\, \mathcal{M}, \label{eq:ADMMassBound}
\end{equation}
which guarantees $\mathcal{A}^\text{max}|_{\theta = \frac{\pi}{2}} \gg \alpha'^{-1}$. This bound remains nonrestrictive in practice, as the extra-dimensional radius $R_\psi$, though much larger than the string length, is still phenomenologically regarded as extremely small.\footnote{Current experimental constraints on the size of extra compact dimensions are relatively weak, requiring only that they be smaller than approximately $\sim 1\,\mu \text{m}$ \cite{Appelquist:2000nn}.}

It is worth noting that we would have reached a qualitatively different conclusion if the tidal stress had remained large but scaled differently with $\epsilon$. For instance, if $\mathcal{A}^\text{max} \sim \mathcal{M}^{-2} \epsilon^{-n}$, the condition for string-scale tidal stress would become $R_\psi^{\frac{2n}{3}} < \alpha' \mathcal{M}^{\frac{2(n-3)}{3}}$, which cannot be satisfied for $0\leq n \leq 3$.

Therefore, the large tidal stress exhibits precisely the right dependence on the KK scale to reach extreme string-scale values when the ADM mass exceeds the bound in \eqref{eq:ADMMassBound}. This significant enhancement compared to the Schwarzschild black hole suggests the onset of stringy transitions in massive probes, potentially leading to their entrapment in the geometry; an effect we investigate in detail in the next section. \\

The equatorial plane is no exception, and the same physics applies to probes released from the equatorial region, $\theta_0 \in [\pi/3, 2\pi/3]$. Numerical solutions to the Hamilton–Jacobi equations \eqref{eq:HamilEqs} confirm that the tidal stress reaches the magnitude derived in \eqref{eq:EquatStressBS} for all probes released from the equatorial region. This behavior is consistent with the ZV geometry, where tidal forces in the equatorial region diverge.

By contrast, probes released from the polar region exhibit different behavior. Numerical solutions show that the tidal stress does not reach the magnitude of \eqref{eq:EquatStressBS} and instead varies significantly with the release angle (see Fig.~\ref{fig:TidForcesComb}). As discussed in Section~\ref{sec:TrajBS}, these probes fall directly into the extremal black holes located at the poles of the bubble. Therefore, the tidal forces they experience are governed by the geometry near these extremal black holes. \\

As shown in \cite{Heidmann:2023kry}, the extremal black holes at the poles are Strominger–Vafa black holes \cite{Strominger:1996sh}, and the near-horizon region is a nontrivial S$^3$ fibration over an AdS$_2 \times$ S$^1$ spacetime: 
\begin{equation}
ds^2_\text{BH} \= d\widetilde{y}^2 + \frac{M(\ell+M)\Delta^2 }{2\ell^4 } \left[ ds(\text{AdS}_2)^2 +4 d\widetilde{\Omega}_3^2\right]
\label{eq:MetricBH}
\end{equation}
where we have introduced $\Delta \equiv \ell^2-M^2 \sin^2\widetilde{\theta} $,  $\widetilde{y}=y-\sqrt{\frac{\ell-M}{\ell+M}} \,t$,   $ds(\text{AdS}_2)^2 = \frac{dR^2}{R^2}-R^2 dt^2$,  and $d\widetilde{\Omega}_3^2$ denotes the metric on a stretched three-sphere:
$$d\widetilde{\Omega}_3^2 =d\widetilde{\theta}^2 + \frac{\ell^6}{\Delta^3} \left(\sin^2 \widetilde{\theta} \,d\phi^2 +\frac{\cos^2\widetilde{\theta}}{4\ell^2} \,d\psi^2 \right).$$
As shown in Fig.~\ref{fig:ProbeRegions}, the infalling probes in the polar region predominantly approach the black hole along its poles at $\widetilde{\theta} = 0$ or $\pi/2$. Thus, the maximal tidal force in the polar region can be estimated by evaluating it using the local geometry \eqref{eq:MetricBH}: 
\begin{equation}
    \mathcal{A}^\text{max}|_{\widetilde{\theta}=0} \sim \frac{\sqrt{3}}{\sqrt{2}\mathcal{M}^2}\,,\qquad \mathcal{A}^\text{max}|_{\widetilde{\theta}=\frac{\pi}{2}} \sim \frac{1}{\sqrt{2}\mathcal{M}^2 \epsilon^3}\,,
    \label{eq:TidalForceSV}
\end{equation}
These results are consistent with the numerical values shown in Fig.~\ref{fig:TidForcesComb}. Probes released near $\theta_0 \sim 0$ (North pole) experience a maximal tidal force consistent with $\mathcal{A}^\text{max}|_{\widetilde{\theta}=0}$, which is slightly stronger than in the Schwarzschild case. Increasing the release angle toward $\theta_0 \sim \pi/3$, the probes begin to fall into the black hole from the South pole (i.e., bubble side), where the tidal force matches $\mathcal{A}^\text{max}|_{\widetilde{\theta}=\frac{\pi}{2}}$. As probes transition into the equatorial region and first interact with the bubble, the tidal force increases by two additional orders of magnitude in $\epsilon$, aligning with \eqref{eq:EquatStressBS}. This microscopic analysis at the cap explains the wide range of maximal tidal forces observed in the ZV and bound-state spacetimes, depending on the release angle, in contrast to the constant distribution for Schwarzschild. \\

The reason some probes in the polar region avoid large tidal forces lies in the presence of extremal black holes at the bubble. Alternative outcomes should arise if one considers more intricate, fully smooth, horizonless geometries: either resolving the extremal black holes into their supersymmetric microstates, such as superstrata \cite{Bena:2015bea,*Bena:2017geu,*Bena:2017xbt,*Heidmann:2019zws,*Heidmann:2019xrd} (for which the complete bound-state background is not known), or into small charged KK bubbles, whose full geometry is known \cite{Bah:2022yji,Bah:2023ows}.
 Although a direct computation of the tidal force in the geometries of \cite{Bah:2022yji,Bah:2023ows} would be very tedious, we can estimate it by using known results about tidal forces in superstratum geometries \cite{Tyukov:2017uig}.

In \cite{Tyukov:2017uig}, it was shown that a probe at the horizon of a Strominger–Vafa black hole experiences a tidal force of magnitude $|Q|^{-1}$, where $Q$ is the black hole charge.\footnote{This expression has the correct dimensionality since the Strominger–Vafa black hole is a five-dimensional black hole. In five dimensions, the charge $Q$ has units [Length]$^2$, yielding a tidal force with units [Length]$^{-2}$.} For our oppositely charged black holes, the charge is given by $|Q| = R_\psi M \sqrt{\frac{\ell+M}{\ell-M}}$, leading to a tidal force of the order of $\epsilon^{-1} \mathcal{M}^{-2}$ if the black holes were isolated. Though the presence of deformation and interaction modifies this result, the estimate remains consistent with our computed values in \eqref{eq:TidalForceSV}, lying roughly midway between the two values.

Moreover, the authors of \cite{Tyukov:2017uig} analyzed tidal forces in superstratum backgrounds: horizonless spacetimes that are coherent microstates of the Strominger–Vafa black hole but replace the horizon by a smooth cap. Despite minimal geometric differences, tidal forces increase dramatically from $|Q|^{-1}$ (black hole) to $|Q|^3 / l_s^8$ (superstratum). Applying this result to our setup suggests that entirely smooth Schwarzschild-like geometries, where the extremal bound states are smoothly capped, should exhibit a similar enhancement. Specifically, we expect an enhancement by a factor $|Q|^4/l_s^8 \sim \frac{\mathcal{M}^8 \epsilon^4}{l_s^8} \gg 1$, indicating that tidal forces in the polar region would be significantly amplified and become string scale as the tidal forces in the equatorial region. \\

To summarize, the bound state of extremal black holes induces tidal forces comparable to those of the Schwarzschild black hole for probes up to the photon orbit. Beyond this region, the replacement of the Schwarzschild horizon by a smooth topological cap leads to a substantial increase in tidal forces. Ultimately, close to the bubble locus, they reach extreme values depending on the KK scale and the ADM mass such that they are above the string scale as long as the mass bound \eqref{eq:ADMMassBound} is satisfied. Although some probes experience weaker tidal forces, this is attributable to the presence of extremal black holes within the bound state. We argued that in fully smooth Schwarzschild-like geometries, tidal forces should be string-scaled at $r\sim 2\mathcal{M}$ for all release angles.

This substantial enhancement relative to the Schwarzschild black hole implies that smooth horizonless geometries may cause tidal disruption in extended probes with internal degrees of freedom. For strings, the string-scaled tidal forces should trigger a stringy transition and result in effective trapping, and this without the presence of a horizon.

\section{Tidal disruption of massless strings}
\label{sec:TidalString}

In this section, we derive the effects of the extreme tidal forces at $r\sim 2\mathcal{M}$ on extended probes possessing internal degrees of freedom, and demonstrate how these forces result in the complete trapping of such entities within this region. To achieve this, we use string probes and determine whether the tidal forces can generate excitations within the internal modes of the string. These excitations will drain the kinetic energy of the infalling strings, thereby preventing their escape. Our analysis will closely follow the derivation conducted for supersymmetric microstate geometries of five-dimensional black holes in \cite{Martinec:2020cml}.

The analysis of string probe dynamics is performed by considering the Penrose limit of the metric \cite{Blau:2002mw}, in which string propagation is solvable in a light-cone gauge. In this limit, the worldsheet dynamics of the transverse coordinates along the string reduce to a set of free fields coupled via a time-dependent mass matrix, which we refer to as the string tidal tensor \cite{Horowitz:1990sr}. We will first review these steps in general before applying them to our three backgrounds of interest: Schwarzschild, Zipoy-Vorhees, and the bound state of extremal black holes. 

\subsection{String propagation in the Penrose limit}

\subsubsection{Penrose limit}
We consider a generic metric with coordinates $(t,r,x^i)$, where $x^i$ includes $\theta$ and  other $x^i$ coordinates associated with Killing vectors:
\begin{equation}
    ds^2 = g_{tt}dt^2 + 2g_{t i}dt dx^i + g_{rr}dr^2 + g_{ij}dx^idx^j.
    \label{eq:metrGENER}
\end{equation}
The Penrose limit first consists of changing coordinates using the affine parameter $\lambda$ of a null geodesic. The equations are given by \eqref{eq:constOfMotion} with $m=0$:\footnote{Note that we normalize the affine parameter in the probe-energy units $p_t=-1$.}
\begin{equation}
    \begin{split}
        \dot{t} &= - g^{tt}+g^{ti}p_i,\qquad \dot{x}^i=g^{ij}p_j-g^{it}, \\
        \dot{r} &= \sqrt{\frac{-g^{tt}+2g^{ti}p_i-g^{ij}p_ip_j}{g_{rr}}},
    \end{split}\label{eq:GeoEqMassless}
\end{equation}
where $p_i$ is the conjugate momentum along $x^i$ \eqref{eq:constOfMotion}. Then, we introduce the following change of coordinates:
\begin{equation}
\left(d t, dr, d x^i\right) \rightarrow\left(\dot{t}, \dot{r}, \dot{x}^i\right) d \lambda+\left(-\frac{d t}{2} , 0, d x^i\right),
\end{equation}
which leads to:
\begin{equation}
    ds^2\= d\lambda dt +2p_i d\lambda dx^i +\frac{g_{tt} \, dt^2}{4} + g_{ij} dx^i dx^j.
\end{equation}
For the Penrose limit to exist, one needs to set $p_i =0$. Since $x^i$ corresponds to a spacetime isometry, except for $\theta$, one can set $p_i=0$ for those coordinates, thereby corresponding to trajectories without momentum along these directions. The case of $p_\theta$ is more subtle. 

For backgrounds admitting separable geodesic equations, like Schwarzschild or the supersymmetric microstate geometries of \cite{Martinec:2020cml}, one can simply take $p_\theta=0$. However, for the ZV or the bound-state geometries, we must restrict to infalling probes in the equatorial plane as we have done before, $\theta = \frac{\pi}{2}$, where $p_\theta=0$. Thus, for those geometries, we will consider string dynamics restricted to this plane, for which the induced metric gives
\begin{equation}
    ds^2 = d\lambda dt +\frac{g_{tt}dt^2}{4}+g_{ij}dx^idx^j.\label{eq:TransformedMetric}  
\end{equation}
Following \cite{Blau:2002mw}, we apply the relativistic scaling that generates the Penrose limit:
\begin{equation}
g_{\mu \nu} \rightarrow \Omega^{-2} g_{\mu \nu} ; \quad \lambda \rightarrow \lambda, \quad t \rightarrow \Omega^2 t, \quad x^i \rightarrow \Omega x^i,
\end{equation}
with $\Omega \to 0$, which yields $d s^2 = d\lambda dt + g_{ij} dx^i dx^j$.
We rewrite the metric in the Brinkman form, a convenient coordinate system for quantizing the string spectrum:
\begin{equation}
    d s^2=2 d x^{+} d x^{-}+\mathcal{A}_{i j}\, z^i z^j {d x^{-}}^2+ d z^i d z^i,
    \label{eq:BrinkmMetr}
\end{equation}
where
\begin{equation}
    \lambda=2 x^{-}, \quad t  = x^{+} - \frac{g_{ij}}{4} \left(\mathcal{Q}^i{ }_k \mathcal{Q}^j{ }_{\ell}\right)' z^i z^j,  \quad x^i= \mathcal{Q}^i{ }_j z^j,\nonumber
\end{equation}
where the prime indicates the derivative with respect to $x^-$. The matrices $\mathcal{Q}$ and $\mathcal{A}$ are defined according to $g$ such as
\begin{align}
    &g_{ij} \mathcal{Q}^i{ }_k \mathcal{Q}^j{ }_{\ell}=\delta_{k \ell},\\
    &g_{ij}\left(\mathcal{Q}^{\prime i}{ }_k \mathcal{Q}^j{ }_{\ell}-\mathcal{Q}^{\prime i}{ }_{\ell} \mathcal{Q}^j{ }_k\right)=0,
    \label{eq:Q&M2} \\
    & \mathcal{A}_{i j}= - (g_{k \ell} \mathcal{Q}^{\prime \ell}{ }_j)' \mathcal{Q}^k{ }_i .
    \label{eq:Aij-Gener}
\end{align}
The matrices $g$, $\mathcal{Q}$ and $\mathcal{A}$ depend only on $x^-$ through $r(x^-)$. The tensor $\mathcal{A}$ plays a role analogous to that of the tidal tensor for massive particles in (\ref{eq:EqOfGeo}), and we will call it the \emph{string tidal tensor}.

\subsubsection{String dynamics}

The bosonic part of the string worldsheet action is given by the Polyakov action:
\begin{equation}
    S_P(X, h)=\frac{1}{4 \pi \alpha^{\prime}} \int d \tau d \sigma \sqrt{h} h^{a b} g_{\mu \nu}(X) \partial_a X^\mu \partial_b X^\nu, \nn
\end{equation}
where the indices $(a,b)$ run over the values $(\tau,\sigma)$, $\alpha'=l_s^2$ is the square of the string length, $h_{ab}$ is the string worldsheet metric, and $h = \det{h_{ab}}$. The fields $X_\mu(\tau, \sigma)$ are the coordinates on the manifold that define embedding into the target space. In Brinkman coordinates, they are given by $X^\mu = (x^+,x^-,z^i)$. 

Using Weyl redundancies, we impose the conformal gauge condition on the worldsheet metric $\sqrt{h}h^{a b}=\eta^{ab}$. In Brinkman coordinates, the action becomes:
\begin{equation}
\begin{aligned}
S_P=\frac{1}{4 \pi \alpha^{\prime}} \int d \tau d \sigma\left[\partial_a x^{+} \partial^a x^{-}+\partial_a z^i \partial^a z^i\right. \\
\left.+\mathcal{A}_{ij} z^i z^j \partial_a x^{-} \partial^a x^{-}\right]. \label{eq:PolyakovBrinkman}
\end{aligned}
\end{equation}
Varying the action with respect to $x^+$ yields the string equations for $x^-(\sigma, \tau)$: 
\begin{equation}
    \left(\partial_\tau^2 - \partial_\sigma^2\right) x^- = 0.
\end{equation}
As shown in \cite{Horowitz:1989bv, Horowitz:1990sr}, this can be solved for pp-wave geometries, for which our solutions are special cases, by the light-cone gauge: $x^{-}=\alpha^{\prime} E \tau$ where $E$ corresponds to the string's center of mass energy. 

The transverse string mode equations yield
\begin{equation}
    \left(\partial_\tau^2-\partial_\sigma^2\right) z^i=(\alpha'E)^2 \mathcal{A}_{ij}\left( \tau\right) z^j.
    \label{eq:StringEqTransv}
\end{equation}
Expanding in Fourier modes, $z^i(\tau, \sigma)=\sum_k z_k^i(\tau) e^{i k \sigma}$, one obtains decoupled harmonic oscillator equations for individual modes:
\begin{equation}
    \ddot{z}_k^i = \left[k^2\delta_{ij}-(\alpha'E)^2 \mathcal{A}_{ij}\right] z_k^j.
    \label{Eq:EOMz}
\end{equation}
Finally, the equation for $x^+$ simply relates $x^+$ to $z^i$, so that the transverse modes $z^i$ capture the string internal degrees of freedom. \\

To analyze the string oscillation in the transverse directions and detect the existence of unstable modes, we solve the equation of motion (\ref{Eq:EOMz}) using the WKB approximation: 
\begin{equation}
    z_k^i \approx \exp \left[ \pm i \int^\tau d \tau \sqrt{k^2 -\left(\alpha^{\prime} E\right)^2 \mathcal{A}_{ii}}\right]. 
\end{equation}
where we have assumed without restriction that $\mathcal{A}$ is a diagonal matrix.\footnote{In all the geometries considered in this paper, $\mathcal{A}$ will be directly diagonal. However, if it happens not to be the case, we can freely perform a change of basis for the $z^i$ in which $\mathcal{A}$ is diagonal.} 
It is more convenient to rewrite this expression in terms of $r$ using the geodesic equation \eqref{eq:GeoEqMassless}, which in terms of $\tau$ gives
\begin{equation}
     \left(\frac{dr}{d\tau}\right)^2 =-4(\alpha'E)^2 \frac{g^{tt}}{g_{rr}}.
     \label{eq:drho/dtau}
\end{equation}
This gives
\begin{equation}
    z_k^i = \exp \left[ \pm \frac{i}{2} \int^r d r \left(-\frac{g_{rr}}{g^{tt}}\right)^{\frac{1}{2}} \sqrt{\frac{k^2}{(\alpha^{\prime} E)^2}- \mathcal{A}_{ii}}\right]. \label{eq:WKBModes}
\end{equation}
Thus, the transverse modes are oscillating in regions where the tidal stress remain mild, $\mathcal{A}_{ii} < \frac{k^2}{(\alpha' E)^2}$, and exhibit an exponential growth and potentially an instability in regions of intense tidal stress $\mathcal{A}_{ii} > \frac{k^2}{(\alpha' E)^2}$. 

The presence of an instability depends on the existence of a turning point $r^*_k$ such that:
\begin{equation}
    \mathcal{A}_{ii}(r^*_k) = \frac{k^2}{(\alpha'E)^2},
    \label{eq:UnstabCondit}
\end{equation}
Finally, we define the following critical threshold value
\begin{equation}
    a_c = (\alpha'E)^{-2},
    \label{eq:muc}
\end{equation}
which is the value at which the string tidal tensor causes exponential growth in the lowest string mode $k=1$. For low-energy strings $\alpha' E \sim \sqrt{\alpha'}$ to be affected by tidal stretching, the critical threshold is string scale, meaning the tidal string tensor must reach string-scale values.

\subsection{In the Schwarzschild geometry}

We first apply the procedure to the Schwarzschild black hole. For such a background, we have $x^i = \{\theta,\phi\}$ and the $g_{ij}$ matrix yields
\begin{equation}
g \= \left(\begin{array}{cccc}
r^2 & 0\\
0 & r^2 \sin^2\theta
\end{array}\right).
\end{equation}
Then, solutions of equations \eqref{eq:Q&M2} and \eqref{eq:Aij-Gener} give
\begin{equation}
    \mathcal{Q} \= g^{-\frac{1}{2}}\,,\qquad \mathcal{A}=  g^{-\frac{1}{2}} (g^{\frac{1}{2}})''.
    \label{matrix form of Q}
\end{equation}
For null geodesics with $p_\theta=0$ in the Schwarzschild metric, one has $r''=\theta'=0$, which implies $$ \mathcal{A}\=0.$$
We recover the known result that transverse string modes in the Schwarzschild black hole do not experience tidal stresses and are freely oscillatory as in Minkowski. 

\subsection{In the Zipoy-Voorhees geometry}

We now consider string dynamics in the ZV spacetime. Since the geometry is not spherically symmetric, we can impose $p_\theta = 0$ only for infalling strings confined to the equatorial plane. We therefore restrict to the plane $\theta = \frac{\pi}{2}$, where the induced metric simplifies to: 
\begin{equation}
    \begin{aligned}
    \left.d     s_{\mathrm{ZV}}^2\right|_{\frac{\pi}{2}} =&\,-\frac{\left(h_{\mathrm{ZV}}-\mathcal{M}\right)^2}{\left(h_{\mathrm{ZV}}+\mathcal{M}\right)^2} d t^2+\frac{\left(h_{\mathrm{ZV}}+\mathcal{M}\right)^3}{4\left(h_{\mathrm{ZV}}-\mathcal{M}\right)} d \phi^2 \\
    & +\frac{4\left(h_{\mathrm{ZV}}-\mathcal{M}\right)\left(h_{\mathrm{ZV}}+\mathcal{M}\right)^5(r-\mathcal{M})^2}{h_{\mathrm{ZV}}^8} d r^2, \nn
\end{aligned}
\end{equation}
where $h_{\mathrm{ZV}}$ is defined in \eqref{eq:GeoZV}. Comparing with the general form \eqref{eq:metrGENER}, there is a single transverse spatial direction, $x^1 = \phi$, and the tensors $\mathcal{Q}$ \eqref{eq:Q&M2} and $\mathcal{A}$ \eqref{eq:Aij-Gener} reduce to scalars. We find:
\begin{equation}
    \mathcal{Q} = \frac{2\sqrt{h_{\mathrm{ZV}}-M}}{\left(h_{\mathrm{ZV}}+M\right)^\frac{3}{2}}, \quad \mathcal{A} = \frac{3 \mathcal{M}^3}{32}\left(\frac{h_{\mathrm{ZV}}}{r(r-2 \mathcal{M})}\right)^5. \label{eq:AStringZV}
\end{equation}
In contrast to Schwarzschild, the tidal potential $\mathcal{A}$ here increases monotonically as the string approaches the singular horizon at $r = 2\mathcal{M}$. Therefore, just as for massive infalling particles, strings experience diverging tidal stresses as they fall deeper into the geometry.

This unbounded growth of $\mathcal{A}$ implies that all transverse string modes governed by \eqref{Eq:EOMz} eventually become unstable: each mode transitions from oscillatory to exponentially growing at the critical radius, $r_k^*$ \eqref{eq:UnstabCondit}, where $\mathcal{A}=\frac{k^2}{(\alpha' E)^2}$. We obtain:
\begin{equation}
\begin{aligned}
r^*_k & =\mathcal{M}\left(1+\sqrt{1+\frac{\gamma_k}{2}\left(\gamma_k+\sqrt{1+\gamma_k^2}\right)}\right),
\end{aligned}
\end{equation} 
where we introduced the dimensionless parameter
\begin{equation}
    \gamma_k = \left(\frac{ \sqrt{3}\alpha'E}{ k\mathcal{M}}\right)^{\frac{2}{5}}.
\end{equation}
The first unstable mode ($k=1$) thus begins to grow at $r = r_1^*$, with higher modes becoming unstable slightly deeper in. For the probe approximation to be valid, the string energy needs to be much smaller than the ADM mass of the background such that $\gamma_k \ll 1$. Moreover, for low-energy strings $\alpha'E\sim \sqrt{\alpha'}$, $\gamma_k$ is even string scale. As a result, the instability onset occurs very close to the singular horizon:
\begin{equation}
2\mathcal{M}\,<\, r \,\leq\, r^*_1 \sim 2 \mathcal{M}\left(1+\frac{\gamma_1}{8}\right).
\end{equation}
Thus, although the tidal stresses on infalling massive particles or massless strings increase only gradually as they approach the singular horizon at $r=2\mathcal{M}$, the tidal disruption of strings erupts abruptly at a string-scale distance from the horizon. This shows that the onset of genuinely new physics — namely, the string instability — occurs only at the horizon scale.

Nevertheless, the appearance of string instability in the ZV geometry is not entirely surprising, since the $r=2\mathcal{M}$ surface is singular and the tidal force diverges there. It was therefore expected that string modes will undergo divergent exponential growth near this surface up until it reaches the singularity. In the next section, we will examine how this result is modified for the bound state of extremal black holes. This configuration is indistinguishable from the ZV geometry up to an infinitesimal distance away from the horizon set by the KK scale. However, it remains smooth at $r = r_+$ on the equatorial plane. A careful analysis is therefore required to determine whether the string instability persists, and whether it remains strong enough to trigger tidal disruption and potentially trap the string within the geometry.

\subsection{In the bound state of extremal black holes}

\begin{figure}[t]
    \begin{center}
    \includegraphics[width=0.483\textwidth]{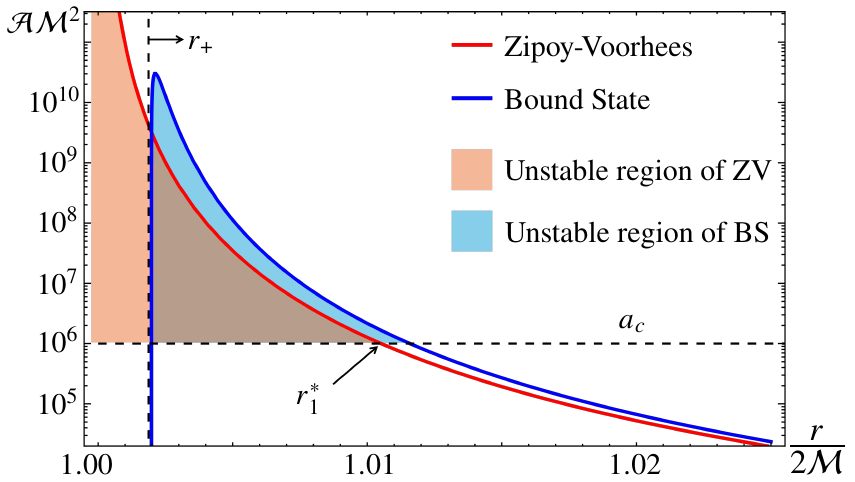}
    \caption{\textit{String tidal stress, $\mathcal{A}_{11}$, of a ZV spacetime and bound-state geometry with $\epsilon = 10^{-2}$ and  the same ADM mass. The threshold $a_c$ \eqref{eq:muc} is arbitrarily chosen, $a_c =10^6 \mathcal{M}^{-2}$.}}
    \label{fig:StringTitdalForces.jpg}
    \end{center}
\end{figure}

We now derive the dynamics of an infalling string in the bound state of extremal black holes on the equatorial plane. The $\theta = \frac{\pi}{2}$ slice of the metric is given by
\begin{align}
    d s_{\mathrm{BS}}^2|_{\frac{\pi}{2}} \=& \medmath{-\frac{(h_\text{BS}-M)^2}{(h_\text{BS}+M)^2}dt^2+dy^2 + \frac{(h_\text{BS}-\ell) (h_\text{BS}+M)}{(h_\text{BS}+\ell) (h_\text{BS}-M)} d\psi^2} \nonumber \\
    & \medmath{+\frac{(h_\text{BS}+\ell)^2 (h_\text{BS}+M)}{ 4(h_\text{BS}-M)} d\phi^2} \\
    &\medmath{+ \frac{4(h_\text{BS}-M)^2(h_\text{BS}+\ell)(h_\text{BS}+M)^4(r-\ell)^2}{h_\text{BS}^8(h_\text{BS}-\ell)}dr^2,} \nonumber
\end{align}
where $h_{\text{BS}}$ is defined in (\ref{eq:GeoBS}). This geometry has three transverse directions, $(x^1,x^2,x^3) = \left(\phi,\frac{\psi}{R_\psi},\frac{y}{R_y}\right)$, with a diagonal metric $g_{ij}$ along these directions.

\begin{widetext}
Solving equations \eqref{eq:Aij-Gener}, we find $\mathcal{A}_{33}=0$ and
\begin{align}
    \mathcal{A}_{22}&\=\medmath{-\frac{16 (\ell - M) h_{\text{BS}}^5 \left[ 2 M h_{\text{BS}}^2 (h_{\text{BS}}+\ell )^2 + \ell M^3 (h_{\text{BS}}+3\ell) + M^2 h_{\text{BS}} (h_{\text{BS}}+2\ell )(3h_{\text{BS}}+\ell) + 2 h_{\text{BS}}^3 (h_{\text{BS}}^2+\ell M ) \right]}{( h_{\text{BS}}+\ell)^3 (h_{\text{BS}} - M)^6 (h_{\text{BS}} + M)^4}},\\
    \mathcal{A}_{11} &\=\medmath{\frac{16 h_{\text{BS}}^5 \left[ (h_{\text{BS}}+M ) \left( 
3 M^3 \left( 2 h_{\text{BS}}^2- (2\ell - M)(\ell + M)  \right) + (\ell - M) h_{\text{BS}} \left(  h_{\text{BS}}^3 - (\ell - M) M (2\ell + 3M)  \right) \right)-3 (\ell - M)^2 M^3 (\ell + M)  \right]}{(h_{\text{BS}}+\ell)^3 (h_{\text{BS}} - M)^6 (h_{\text{BS}} + M)^4}}. \nonumber
\end{align}
\end{widetext}
First, one can check that we recover the ZV result \eqref{eq:AStringZV} when $\ell = M = \mathcal{M}$, in which case $\mathcal{A}_{22} = \mathcal{A}_{33} = 0$ and $\mathcal{A}_{11} = \mathcal{A}$, as expected. Thus, for macroscopic bound states satisfying \eqref{Eq:l&M@EPSorder}, the string tidal tensor matches that of ZV spacetime up to very close to the cap, beyond which it starts to differ.

We observe that $\mathcal{A}_{22}$ is always negative, ranging from order $-\epsilon$ just outside the bubble locus to $-\frac{112}{\mathcal{M}^2 \epsilon^5}$ exactly at the locus. Consequently, string oscillations along the $\psi$ direction remain stable, undergoing free oscillation during infall with a varying frequency.

In contrast, instabilities may arise in the transverse modes along the $\phi$ direction. These modes follow the ZV result closely until near the cap. Analyzing $\mathcal{A}_{11}$, we find that it agrees with the ZV value up to a range of order $\epsilon \mathcal{M}$ above the would-be horizon as expected. There, it reaches a maximum before rapidly falling to a large negative finite value of order $-\frac{16}{\mathcal{M}^2 \epsilon^5}$. We illustrate this behavior for a bound state with $\epsilon = 10^{-2}$ in Fig.~\ref{fig:StringTitdalForces.jpg}, together with the corresponding result for the ZV geometry.

Expanding near the cap, the maximum tidal stress $\mathcal{A}^\text{max}$ and its location $r_\text{max}$ are given by
\begin{equation}
    r_\text{max}= r_+ + \frac{\epsilon\mathcal{M}}{20}=2\mathcal{M}\left(1+\frac{17\epsilon}{80} \right), \quad \mathcal{A}^\text{max}=\frac{3}{ \mathcal{M}^2 \epsilon^5}.
\end{equation}
Thus, the string undergoes intense tidal stress close to the cap, around $r \sim 2\mathcal{M}$, with the same functional dependence on the KK scale as the tidal stress experienced by a massive infalling particle \eqref{eq:EquatStressBS}. Remarkably, in the immediate vicinity of the bubble, the tidal stress becomes negative again, showing that nothing singular occurs at the smooth bubble locus and that the string is freely oscillating there. Therefore, if a region of string instability exists, i.e. where $\frac{k^2}{(\alpha' E)^2}-\mathcal{A}_{11}<0$, it must lie in a narrow shell slightly above the bubble. 

Such a region exists only if $\mathcal{A}^\text{max} > a_c = (\alpha' E)^{-2}$, which gives a condition between the string energy, the KK scale, the string scale, and the ADM mass:
\begin{equation}
    \mathcal{M}^2 \epsilon^5 < (\alpha' E)^2 \quad \Leftrightarrow\quad \frac{R_\psi^5}{\mathcal{M}^2} < (\alpha' E)^3, \label{eq:StringInstBound2}
\end{equation}
where we have dropped irrelevant numerical prefactors. 

This inequality implies that the string must have a minimum energy in order for tidal effects to induce an instability. Below this energy threshold, tidal forces are too weak to excite transverse modes, and the string remains unaffected.

As argued in the previous section, the tidal stress reaches extreme string-scale values as long as the ADM satisfies the bound \eqref{eq:ADMMassBound}, which implies $\frac{R_\psi^5}{\mathcal{M}^2} < \frac{\alpha'^2}{R_\psi} \ll \alpha'^{3/2}$. Thus, for those bound states, we are guaranteed that low-energy strings with $\alpha' E \sim \sqrt{\alpha'}$ satisfy \eqref{eq:StringInstBound2} and are affected by the tidal stretching.

Moreover, one can use the argument detailed in the paragraph below \eqref{eq:ADMMassBound} to show that the region of exponential growth is strongly localized near the bubble locus. Indeed, we are guaranteed that $\frac{k^2}{(\alpha' E)^2} - \mathcal{A}_{11} > 0$ for $\alpha' E \sim \sqrt{\alpha'}$ as soon as $\mathcal{A}_{11} \lesssim \mathcal{M}^{-2} \epsilon^{-3}$, which occurs for $r \gtrsim 2\mathcal{M}(1 + \epsilon^{3/5})$. Thus, above that radius, the string modes are necessarily freely oscillating.\\

Assuming the condition \eqref{eq:ADMMassBound} is met, an infalling string with energy above the bound \eqref{eq:StringInstBound2} experiences exponential growth in a small region near the cap. This occurs for the transverse modes with $1 \leq k \leq k_\text{max}$, where the highest unstable mode number $k_\text{max}$ is
\begin{equation}
    k_\text{max} \equi \frac{\sqrt{3}\,\alpha' E}{\mathcal{M} \epsilon^\frac{5}{2}}.
\end{equation} \\

Thus, we obtain a result that closely parallels the analysis of supersymmetric smooth horizonless geometries in \cite{Martinec:2020cml}, now realized in a realistic Schwarzschild-like background. As in \cite{Martinec:2020cml}, we find a region where certain oscillator modes of the infalling string experience exponential growth. This region is not located precisely at the cap but lies slightly above it and extends outward. However, in contrast to the solutions of \cite{Martinec:2020cml}, where the instability stretches high within the black-hole throat, in our case it remains confined to the near-horizon region of the Schwarzschild black hole, $r \sim 2\mathcal{M}$, where the classical horizon is replaced by the topological structure. \\

However, unlike in the ZV geometry, the region of exponential growth here is finite. As a result, the string’s oscillator modes grow only for a limited time: once during infall, before reflecting off the bubble locus, and once again on the way out. The instability is therefore qualitatively different from that in ZV spacetime: the transverse modes do not diverge but instead exhibit transient, finite-amplitude growth.

As discussed in \cite{Horowitz:1990sr,Martinec:2020cml}, these excited oscillator modes drain kinetic energy from the infalling string, effectively converting it from a massless to a massive state. Following the analysis of \cite{Horowitz:1990sr,Martinec:2020cml}, we will estimate the mass acquired by the string after traversing the unstable region and determine the maximum height it reaches upon rebound due to this induced mass.

\subsubsection{Induced mass and bouncing height}

We now estimate the average mass acquired by a string after it bounces off the smooth bubble locus, traversing the unstable region twice. Our analysis closely follows that of \cite{Horowitz:1990sr,Martinec:2020cml}, to which we refer the interested reader for further details. We begin by evaluating the average number of excited modes produced, $\langle \mathcal{N}_\text{osc} \rangle$, given in terms of the Bogoliubov coefficient $\beta_k$ as follows:
\begin{equation}
    \langle \mathcal{N}_\text{osc} \rangle \= \sum_k k |\beta_k|^2.
\end{equation}
The Bogoliubov coefficient characterizes the amplitude of the reflected wave relative to the transmitted wave across a potential barrier. In the WKB approximation, this coefficient is nonzero only for $k \leq k_\text{max}$ and is approximately given by the square of the expression \eqref{eq:WKBModes}, with the integral evaluated over the unstable region:
\begin{equation}
    |\beta_k|^2 \sim \exp \left[\int_\text{uns. reg.} d r \left(-\frac{g_{rr}}{g^{tt}}\right)^{\frac{1}{2}} \sqrt{\mathcal{A}_{11}-\frac{k^2}{(\alpha^{\prime} E)^2}}\right]. \nn
\end{equation}
Since $\mathcal{A}_{11}$ is a complicated function of $r$, we can only estimate this integral for small $\epsilon$ and $\alpha' E$, assuming $k_\text{max}>1$:
\begin{equation}
    |\beta_k|^2 \sim \exp \left[ 3 k_\text{max}^\frac{1}{3} \left( (k_\text{max})^{\frac{k_\text{max}^{-\frac{1}{3}}}{3}}-k^{\frac{k_\text{max}^{-\frac{1}{3}}}{3}}\right)\right]. 
\end{equation}
We then approximate the total mode production by converting the sum into an integral:
\begin{equation}
    \langle \mathcal{N}_\text{osc} \rangle \sim c\,k_\text{max}^2,
\end{equation}
where $c$ is an $\mathcal{O}(1)$ constant that approaches unity in the large $k_\text{max}$ limit.

Remarkably, although the tidal stress on the string is extremely large, the mode production remains moderate, behaving as if $|\beta_k|^2 \sim 2$ for $k$ in the range $[1, k_\text{max}]$. This results from the fact that the string traverses the small unstable region in a very short amount of proper time. \\

Once the string exits this region of strong tidal stress, the number of excited modes, $\langle \mathcal{N}_\text{osc} \rangle$, effectively contributes to its mass. In regions of weak tidal stress, the transverse modes can be integrated out in the Polyakov action \eqref{eq:PolyakovBrinkman}, leading to $\int_\sigma \partial_a z^i \partial^a z^i \propto \langle \mathcal{N}_\text{osc} \rangle$ \cite{Horowitz:1990sr,Martinec:2020cml}.

Hence, the string dynamics can be approximated by that of a point particle with mass $m^2 = \alpha' \langle \mathcal{N}_\text{osc} \rangle$ and energy $\alpha' E$. The radial geodesic equation for such a massive particle reads:
\begin{equation}
    \left(\frac{d r}{d \tau}\right)^2=\frac{(\alpha'E)^2}{g_{r r}}\left(-g^{t t}- \frac{\left\langle\mathcal{N}_{\mathrm{osc}}\right\rangle}{\alpha' E^2}\right).
\end{equation}
The maximum radial distance reached, i.e., the turning point, is determined by the zero of $\frac{d r}{d \tau}$. Since $-g^{tt} > 1$, this turning point exists only if $\left\langle\mathcal{N}_{\text{osc}}\right\rangle > \alpha' E^2$. If this is not satisfied, the string retains sufficient energy to escape, despite having lost kinetic energy to the excited modes. We thus define a key dimensionless parameter:
\begin{equation}
    \zeta \equi \sqrt{\frac{\alpha'E^2}{\left\langle\mathcal{N}_{\mathrm{osc}}\right\rangle}} \,\sim\, \frac{\mathcal{M}^2 \epsilon^5}{\alpha'},
    \label{eq:zeta}
\end{equation}
where numerical prefactors have been omitted for clarity. Importantly, $\zeta$ is independent of the string energy. Thus, strings will reach the exact same maximum height independent from their initial energies. This is because higher-energy strings, while they could go higher up, also generate more excited modes draining more energy out, and, remarkably, the two effects exactly cancel.

The trapping condition $\zeta < 1$ coincides with the condition for the bound state to generate string-scale tidal stress, given in \eqref{eq:ADMMassBoundfirst}. Moreover, with the bound \eqref{eq:ADMMassBound} for having extreme string-scale values, we have
\begin{equation}
    \zeta \,<\, \left(\frac{l_s}{R_\psi} \right)^\frac{2}{3} \ll 1\,.
\end{equation}

As before, a very different scenario would have occurred if the tidal stress generated by the geometry had scaled with the KK scale more mildly (as $\mathcal{M}^{-2} \epsilon^{-3}$ or lower). Indeed, $\zeta$ would have been $\sim (R_\psi/l_s)^2$ or larger, exceeding unity, and all strings, even those affected by the tidal instability, would have escaped instead of becoming trapped. \\

Assuming \eqref{eq:ADMMassBound} and $\zeta \ll 1$, the maximal radius reached by the string, $r_\text{max}$, is obtained from the turning point condition $\frac{d r}{d \tau} = 0$:
\begin{equation}
    r_{\text{max}}=\ell+\frac{M\sqrt{1+\zeta^2-\zeta}}{1-\zeta}.
\end{equation}
This exceeds the bubble radius $r_+$ \eqref{eq:Defrpm} only if $\zeta > (\ell - M)/(\ell + M) \sim \epsilon/4$. When $\zeta < \epsilon/4$, the effective mass generated by the excited modes is too large for the string to escape even minimally from the cap. In such cases, the string does not complete a bounce and is trapped after passing through the unstable region only once.

The condition for the string to properly exit the unstable region, $\zeta > \epsilon/4$, translates into
\begin{equation}
    \mathcal{M} \,<\, \left(\frac{R_\psi}{l_s} \right)^4 \,R_\psi\,, \label{eq:MassBound}
\end{equation}
and the maximum height reached by the string is
\begin{equation}
    r_\text{max} \= r_+ \left(1 + \frac{\zeta}{4}- \frac{\epsilon}{16} \right) \sim 2\mathcal{M}\left(1 + \frac{\zeta}{4}- \frac{\epsilon}{16} \right).
\end{equation}
For bound states with ADM mass exceeding \eqref{eq:MassBound}, we have $\zeta < \epsilon/4$, and the string becomes even more trapped, failing to complete a bounce.

Thus, for all bound states satisfying $\mathcal{M} > \left(\frac{R_\psi}{l_s} \right)^2 R_\psi$, strings are guaranteed to become trapped near the cap, that is, the near-horizon geometry of the Schwarzschild black hole at $$ r \,\lesssim\, 2\mathcal{M}\left(1+\mathcal{O}\left(\epsilon,\zeta\right)\right).$$ Remarkably, this happens independently of their initial energy, provided \eqref{eq:StringInstBound2} holds.

Overall, our result mirrors the behavior observed in supersymmetric microstate geometries \cite{Martinec:2020cml}, but now within the more realistic Schwarzschild regime. However, while in \cite{Martinec:2020cml} the strings typically rebounded far from the cap, in our case the entire dynamical process is strikingly confined to the vicinity of $r \sim 2\mathcal{M}$. This confinement would not have occurred without the bound state's delicate interplay between the KK scale, the string scale, and its ADM mass.  \\

For the ZV geometry, the tidal instability of strings infalling along the equator was evident from exponential divergence of transverse modes close to $r \sim 2\mathcal{M}$. In contrast, in the bound-state geometries, the equator is smooth and horizonless so the growth region is confined to a narrow shell around $r \sim 2\mathcal{M}$ and lasts only a short time. Nevertheless, we find that the maximal tidal stresses have the perfect features leading to a dramatic excitation of internal string modes, causing strings over a wide energy range \eqref{eq:StringInstBound2} to become trapped near $r \sim 2\mathcal{M}$.

The first of these effects lies in the scaling of the maximal tidal stress as $\mathcal{M}^{-2} \epsilon^{-5}$, which ensures strings experience instability in a wide energy range. Had the stress scaled more mildly (e.g., as $\epsilon^{-3}$), the maximal stress would not have reached string-scale values, few strings would have been tidally disrupted, and most would have escaped, yielding a completely opposite scenario.

The second effect lies in the fact that the mass induced by the internal modes scales precisely as $\langle \mathcal{N}_{\mathrm{osc}} \rangle \propto (\alpha' E)^2$, making tidal trapping energy-independent. Strings with sufficient energy to probe the unstable tidal region will experience the same energy drain, keeping them near the cap, regardless of their initial energy. 

Therefore, even though the underlying geometry is free of singularities or horizons on the equatorial plane, the dynamics of infalling strings effectively mimic those in black hole backgrounds: they are absorbed into the region around $r \sim 2\mathcal{M}$ and cannot return. This behavior is a robust feature of the bound state geometry and underlies its capacity to reproduce black hole-like absorption for a wide class of probes with internal degrees of freedom, providing strong support for interpreting these features as the \emph{emergence of an effective horizon}.

\subsubsection{Remarks on untrapped strings and off-equatorial trajectories}

First, although the tidal trapping induced by the geometries is remarkably insensitive to the string energy, it still requires the energy to exceed the bound \eqref{eq:StringInstBound2} so that the internal string modes are excited while passing through the tidal instability region. Hence, massless strings with energy below
\begin{equation}
    \alpha' E \,<\, \frac{R_\psi^\frac{5}{3}}{\mathcal{M}^\frac{2}{3}} \,<\, \left(\frac{\sqrt{\alpha'}}{R_\psi}\right)^\frac{1}{3}\,\sqrt{\alpha'}, \label{eq:EnergyBound}
\end{equation}
will not be trapped and instead reflect off the geometry, where we have used the bound \eqref{eq:ADMMassBound} for the second inequality. This bound restricts to strings with energies well below the string scale. Since there is no lower limit to the energy of a massless string in four-dimensional vacuum, this suggests that although smooth, horizonless Schwarzschild-like geometries can absorb most of the probe energy spectrum, a faint shower of stable, low-energy particles may still escape from the geometry. \\

Second, the analysis focused exclusively on strings falling along the equatorial plane. While this is the only scenario where an analytic analysis is possible, we can qualitatively extend our conclusions to off-equatorial trajectories by extrapolating from previous results.

 In Section \ref{sec:TidalStress}, we have shown that massive probes falling within the equatorial region, $\theta \in [70^\circ,110^\circ]$, experience tidal forces of the same order as those along the equator. We therefore expect a similar trapping effect at the cap for strings falling within this angular range.

For massive probes falling through the polar region, the maximal tidal force is smaller than $\epsilon^{-3} \mathcal{M}^{-2}$, i.e. the critical threshold below which strings with $\alpha' E \sim \sqrt{\alpha'}$ do not experience a tidal instability. However, as discussed in Section \ref{sec:TidalParticleBS}, the weak tidal forces in this region arise because the geometries are only partially smooth, and the probes primarily interact with the extremal black holes in the bound state. By instead considering more intricate geometries in which the Schwarzschild horizon is fully resolved into a smooth, horizonless topology, we expect the tidal forces in the polar region to be significantly enhanced, reaching string-scale values that excite internal string modes and induce trapping.

In conclusion, we expect that strings with energy above the bound \eqref{eq:StringInstBound2}, falling from any direction toward a smooth, horizonless Schwarzschild topological soliton, will undergo a tidal instability exactly at the cap, i.e. the horizon scale $r \sim 2\mathcal{M}$, leading to their sudden trapping and preventing escape.

\section{Conclusion and outlook}
\label{sec:conclusion}

\vspace{-0.2cm}
\subsection*{Conclusion}
\vspace{-0.2cm}

In this work, we have shown that smooth, horizonless geometries in supergravity, closely resembling the Schwarzschild black hole but with the horizon replaced by a highly redshifted, smooth topological cap, can induce horizon-like effects on infalling probes. As a representative prototype of such Schwarzschild-like microstructure, we analyzed the dynamics of point particles and strings falling into the neutral, regular bound state of two extremal black holes constructed in \cite{Heidmann:2023kry}.

Our analysis demonstrated that probes follow Schwarzschild-like radial trajectories down to the photon orbit, $r\sim 3\mathcal{M}$, with only mild deviations below this radius. However, unlike in the Schwarzschild geometry, some geodesics experience rapidly growing tidal forces induced by the smooth topological structure near the would-be horizon. These tidal stresses reach extreme values near the cap and depend only on the KK scale and the ADM mass, attaining extreme string-scale magnitudes under the mild condition $\left(\frac{R_\psi}{l_s}\right)^{2} R_\psi \,<\, \mathcal{M}$. For geodesics that do not encounter such strong tidal effects, we argued that this results from the presence of the two extremal black holes; further resolving them into fully horizonless geometries, via superstrata or the constructions of \cite{Bah:2022yji,Bah:2023ows}, is expected to enhance the tidal forces to string-scale intensities.

When point particles are replaced by extended probes such as massless strings, these extreme tidal forces trigger an instability that mimics horizon-like behavior: the incident kinetic energy is substantially transferred into internal excitations, preventing the probe from escaping after reaching the cap. Instead, the probe becomes trapped, with its turning point located deep within the cap region, around $r_\text{max} = 2\mathcal{M}\left(1+\mathcal{O}(\epsilon,\zeta)\right)$, where $\epsilon$ \eqref{eq:Defepsilon} and $\zeta$ \eqref{eq:zeta} are infinitesimal dimensionless parameters depending on the string length, the ADM mass and the extra dimension radius. 

\vspace{-0.2cm}
\subsection*{Future directions}
\vspace{-0.2cm}

A natural follow-up question arising from the results of this paper concerns the fate of probes trapped in the cap region. In the case of strings, one can expect fragmentation into light closed-string excitations, a fraction of which may be re-emitted as a muffled burst of radiation. For this radiation to escape, however, the energy of the constituent strings must be sufficiently low to satisfy the bound in \eqref{eq:EnergyBound}, ensuring they are not themselves tidally disrupted. Addressing this requires investigating the potentially observable effects of radiation carrying infinitesimal energy from the microstructure. In particular, for gravitons or gravitational waves, this would correspond to ultra-low-frequency, long-wavelength emissions originating from the black hole microstructure. \\

To make the analysis tractable, several simplifying assumptions were adopted in this work. In particular, we focused on the not fully horizonless bound state of extremal black holes, rather than the more intricate smooth solutions of \cite{Bah:2022yji,Bah:2023ows}, and restricted most of the analytical treatment to the equatorial plane, the symmetry plane of the geometry. It would be worthwhile to revisit the main conclusions of this study in the context of fully smooth horizonless solutions,\footnote{We could also be interested in analyzing the neutral bound states of many black holes constructed in \cite{Dulac:2024cso,*Heidmann:2023thn}.} and to extend the analysis to probes falling outside the equatorial plane. \\

More broadly, it would be interesting to go beyond the bosonic string and include the full superstring dynamics, incorporating worldsheet fermions and their couplings to background gauge fields. One expects qualitatively similar behavior, although the gauge fields could potentially extract additional kinetic energy from the center-of-mass motion of the probe.\\

A longer-term goal inspired by this work is to extend the analysis to rotating black holes, where the presence of an ergoregion could substantially modify the dynamics. This would first require constructing smooth horizonless geometries that asymptote to the Kerr solution, in analogy with the Schwarzschild-like configurations studied here. Recent developments in this direction offer promising prospects for achieving this in the near future \cite{Chakraborty:2025ger}.

\acknowledgments 

We thank Nima Arkani-Hamed, Ramy Brustein, Raphaël Dulac, Gary Horowitz, Emil Martinec, Nick Warner and Yoav Zigdon for insightful discussions and Matthias Blau for valuable correspondence and clarifications. The work is supported by the Department of Physics at The Ohio State University.

\bibliography{microstates}

\clearpage \appendix

\end{document}